\documentclass[prd,preprintnumbers,aps,preprint,amsmath,amssymb,showkeys,superscriptaddress,
showpacs,floatfix,nofootinbib,unsortedaddress]{revtex4}

\usepackage{amssymb, bm, amsmath, graphicx, subfigure, physymb}
\usepackage[usenames,dvipsnames]{color}
\usepackage{subfigure}
\usepackage{slashed}
\usepackage{multirow,array}
\usepackage[utf8]{inputenc}
\usepackage{array}
\usepackage{enumerate}
\usepackage{amsfonts}
\usepackage{dsfont}
\usepackage{mathtools}
\usepackage{mathrsfs}
\usepackage{slashed}
\usepackage{latexsym}
\usepackage{hyperref}
\usepackage{epstopdf} 

\pdfoutput=1
\newcommand{\beq}{\begin{equation}}
\newcommand{\eeq}{\end{equation}}
\newcommand{\bqa}{\begin{eqnarray}}
\newcommand{\eqa}{\end{eqnarray}}
\newcommand{\ben}{\begin{eqnarray*}}
\newcommand{\een}{\end{eqnarray*}}
\newcommand{\bes}{\begin{subequations}}
\newcommand{\ees}{\end{subequations}}

\newcommand{\beal}{\begin{align}}

\newcommand{\amu}{\alpha_\mu}

\newcommand{\stackeven}[2]{{{}_{\displaystyle{#1}}\atop\displaystyle{#2}}}

\newcommand{\gsim}{\stackeven{>}{\sim}}

\newcommand{\dhd}{{\textstyle d}
\lower.03ex\hbox{\kern-0.38em$^{\scriptstyle-}$}\kern-0.05em{}}
\newcommand{\dbar}{{\textstyle \delta}
\lower.03ex\hbox{\kern-0.38em$^{\scriptstyle-}$}\kern-0.05em{}}
\newcommand{\half}{{1\over 2}}

\newcommand{\gPM}{g^{+-}}
\newcommand{\gpm}{g_{+-}}
\newcommand{\ubar}[1]{\overline{U}_{#1}}

\newcommand{\Dtwo}[2]{\hat{D}_2 ( \bm #1 , \bm #2 )}
\newcommand{\Dfour}[4]{\hat{D}_4 ( \bm #1 , \bm #2 , \bm #3 , \bm #4 )}
\newcommand{\Dsix}[6]{\hat{D}_6 ( \bm #1 , \bm #2 , \bm #3 , \bm #4, \bm #5 , \bm #6 )}
\newcommand{\Deight}[8]{\hat{D}_8 ( \bm #1 , \bm #2 , \bm #3 , \bm #4, \bm #5 , \bm #6, \bm #7, \bm #8 )}

\newcommand{\oone}{
\begin{picture}(10,8)
\put(5,5){\circle{8}}
\put(2.9,2.5){{\scriptsize 1}}
\end{picture}
}
\newcommand{\otwo}{
\begin{picture}(10,8)
\put(5,5){\circle{8}}
\put(2.9,2.5){{\scriptsize 2}}
\end{picture}
}
\newcommand{\othree}{
\begin{picture}(10,8)
\put(5,5){\circle{8}}
\put(2.9,2.5){{\scriptsize 3}}
\end{picture}
}
\newcommand{\ofour}{
\begin{picture}(10,8)
\put(5,5){\circle{8}}
\put(2.9,2.5){{\scriptsize 4}}
\end{picture}
}

\newcommand{\ord}[1]{\mathcal{O} \left( #1 \right)}
\newcommand{\tr}{\mathrm{tr}}

\begin{document}

\title{Toward Initial Conditions of Conserved Charges\\ Part I: Spatial Correlations of Quarks and Antiquarks}
\author{Mauricio Martinez}
\email[Email: ]{mmarti11@ncsu.edu}
\affiliation{Department of Physics, North Carolina State University, Raleigh, NC 27695, USA}
\author{Matthew D. Sievert}
\email[Email: ]{sievertmd@lanl.gov}
 \affiliation{Theoretical Division, Los Alamos National Laboratory, Los Alamos, NM 87545, USA}
\author{Douglas E. Wertepny}
\email[Email: ]{douglas.wertepny@usc.es}
\affiliation{Departamento de F\'isica de Part\'iculas and IGFAE, Universidade de Santiago de Compostela, 15782
Santiago de Compostela, Galicia-Spain}

\begin{abstract}
In this paper, we study the spatial correlations among quarks and antiquarks produced at mid-rapidity by gluon pair production in the color glass condensate framework.  This paper is the first part of a series in which we calculate a complete set of quark/quark, quark/antiquark, and antiquark/antiquark spatial correlation functions in heavy-light ion collisions, with the goal of incorporating their conserved charges into the initial conditions of hydrodynamics.  The physical mechanisms captured in this calculation include geometric, entanglement, and interaction-mediated correlations.  In this first paper, we construct the building blocks for the correlations arising from single- and double-pair production, studying in detail the single-pair case and the general features of the double-pair case.  We find a rich correlation structure in transverse coordinate space, with different mechanisms dominating over different length scales, and we present explicit results for the quark-antiquark correlations in the single-pair production regime.  We reserve a detailed discussion of the double-pair production regime for the next paper in this sequence.  
\end{abstract}

\date{\today}
\maketitle

%
\section{Introduction}
\label{sec:intro} 
%

In recent years, fluctuations in the initial stages of heavy ion collisions have become understood as central drivers of the overall spectrum of produced particles.  For instance, while the phenomenon of elliptic flow $v_2$ can be understood mainly as an average geometric effect in peripheral heavy ion collisions, the triangular flow $v_3$ seems to be driven entirely by spatial fluctuations in the initial energy density of the fireball~\cite{Alver:2010gr}.  In addition to the nucleon-scale fluctuations arising from the distribution of nucleons within the colliding ions in Monte-Carlo-Glauber models (see for instance Ref.~\cite{Loizides:2014vua} and references therein), sub-nucleonic fluctuations also leave an important imprint on the spectrum of produced hadrons~\cite{Gale:2012rq,Mantysaari:2017cni}.  Unlike the larger-scale fluctuations due to nucleonic degrees of freedom, these sub-femtometer fluctuations due to QCD degrees of freedom are driven largely by color charge.  

In heavy ions at high energies, the Lorentz contraction of multiple nucleons into a short longitudinal thickness enhances the RMS two-dimensional color-charge density.  These parametrically large charge densities radiate intense, classical gluon fields described by the effective theory known as the color glass condensate (CGC) (see \cite{Kovchegov:2012mbw} and references therein).  The sub-nucleonic fluctuations in energy density due to these classical gluon fields are incorporated into, for instance, the IP-Glasma model~\cite{Schenke:2012wb,Schenke:2013dpa}, with the sub-femtometer fluctuations having a substantial effect on particles observed in the final state~\cite{Mantysaari:2017cni,Mantysaari:2016jaz,Mantysaari:2016ykx,Mantysaari:2017dwh}.

Fluctuations and correlations in the initial stages of proton-proton, proton-nucleus, and nucleus-nucleus collisions have been the subject of tremendous activity within the CGC community (see for instance the reviews \cite{Dusling:2015gta} and \cite{Venugopalan:2016vgm} and the references therein).  Much of this work focuses on the applications of the CGC formalism as a natural explanation for the long-range-in-rapidity ``ridge'' correlations in nucleus/nucleus collisions \cite{Dumitru:2008wn,Dusling:2009ni,Schenke:2016ksl} as well as high-multiplicity proton-proton \cite{Dumitru:2010iy,Dusling:2013qoz} and proton/nucleus \cite{Dusling:2012iga,Dusling:2013qoz,Dusling:2015rja} collisions.  Similar long-range-in-rapidity azimuthal modulations of the correlation function are also natural candidates for CGC effects in the initial state \cite{Dumitru:2014yza}.  Other works such as Refs.~\cite{Lappi:2009xa,Schenke:2015aqa} study how such correlations produced in the initial stages of hadronic collisions may be modified by the subsequent strong-coupling classical Yang-Mills dynamics of the gluon fields.  Other approaches to studying the role of initial-state correlations include the AMPT model \cite{Zhang:1999bd,Pang:2015zrq}, the longitudinally-extended source model \cite{Broniowski:2015oif}, and others.

These studies of the initial conditions of heavy ion collisions have so far focused predominantly on the fluctuations of the initial energy density, which is dominated by the production of gluons.  However, new state-of-the-art hydrodynamic codes have been produced which are capable of preserving other conserved charges, such as flavor and baryon number, throughout their evolution~\cite{Shen:2017bsr}.  Moreover, recent {\textit{ab initio}} CGC calculations have computed certain momentum-space correlations between quarks and made strides toward extending the dilute-dense formalism to the dense-dense limit~\cite{Altinoluk:2016vax,Kovner:2017gab,Kovner:2017ssr,Kovner:2018vec,Lappi:2015vta}.  In light of these simultaneous advances, there is a timely opportunity to compute the full set of quark and antiquark correlations in the CGC, along with their associated conserved charges such as flavor, baryon number, and electric charge.  These can then be combined with the latest hydrodynamic simulations to assess their impact on the spectrum of final-state particles.

It has long been known that baryon stopping, the production of {\textit{net}} baryon number at mid rapidity, is suppressed at high energies \cite{Itakura:2003jp,Kharzeev:1996sq,Capella:1996th,Rossi:1977cy}. The same is true of other quantum numbers, like spin and flavor, which are carried by valence quarks.  Most of the valence quantum numbers brought by the colliding ions are carried down the beam pipe or to very far forward rapidities, with the dynamics at mid rapidity being dominated by the soft gluon fields of the CGC.  While these soft gluons themselves carry no net flavor or baryon number, they do produce a spectrum of characteristic {\textit{fluctuations}} through quark-antiquark pair production.  One may attempt to identify the momentum-space correlations among quarks and antiquarks with the interesting correlations~\cite{Adam:2016iwf} observed in experiment among baryons and antibaryons.  Some caution in such an identification is warranted, however, because nonperturbative hadronization effects may muddle the connection implied by parton-hadron duality.  In our case, we are interested primarily in the spatial correlations at the parton level which serve to specify the initial conditions for the energy-momentum tensor and conserved currents in hydrodynamics.  These initial-state effects are free from hadronization corrections, and their impact on final-state particle production can only be determined by coupling these initial conditions to subsequent hydrodynamic evolution.

The production of a single $q \bar q$ pair in proton-nucleus collisions in the CGC has been studied in the past by various authors~\cite{Kovchegov:2006qn,Gelis:2004jp,Blaizot:2004wv,Fujii:2006ab,Gelis:2003vh,Levin:1991ry}.  Although these calculations focused on the production cross-sections in momentum space, the same essential ingredients they derived are responsible for local fluctuations and quark/antiquark correlations in coordinate space.  An important recent calculation by Altinoluk et al.~\cite{Altinoluk:2016vax} examined the effect of Pauli blocking by studying quark/quark correlations via double pair production in heavy ion collisions.  The correlations obtained in~\cite{Altinoluk:2016vax} neglect the contributions in which the radiated gluon first scatters in the target field and later fragments into a $q \bar{q}$ pair.  This contribution was knowingly omitted by the authors~\cite{Altinoluk:2016vax} based on the expectation that gluon scattering followed by pair production will not lead to the Pauli blocking correlations of interest.  However, as we will show explicitly for the case of single-pair production, it is precisely the interferences of these terms which generate the dominant contribution to quark / antiquark correlations.  It is therefore reasonable to expect that contributions of this type may also generate significant interaction-mediated correlations in the quark/quark case, although these are separate from the statistical Pauli blocking effects that were the focus of~\cite{Altinoluk:2016vax}.  We will explore these aspects in detail in our future work.

An important extension of the CGC formalism for dilute-dense collisions to ``semi-dilute--dense'' collisions was recently developed by Kovchegov and Wertepny \cite{Kovchegov:2012nd}.  This formalism, intended to describe asymmetric ``heavy-light ion collisions'' such as copper-gold collisions, completely re-sums the high-density effects in the heavy ion while incorporating the high-density effects of the light ion order-by-order in perturbation theory.  Such effects enhance the probability to radiate a second $q \bar q$ pair, enabling additional types of correlations through quantum entanglement.

Building on these recent advancements, we will undertake the calculation of the leading-order quark/quark, quark/antiquark, and antiquark/antiquark spatial correlations at mid-rapidity, along with their associated charges, in the CGC formalism.  We will first analyze in detail the correlations arising from a single $q \bar q$ pair production at mid-rapidity; this contribution describes the leading-order correlations in proton-nucleus collisions and remains the dominant short-distance contribution in heavy-light ion collisions.  We will also study at length the correlations arising from the production of two $q \bar q$ pairs at mid-rapidity; these correlations, while still sub-leading compared to single-pair production, are enhanced in heavy-light ion collisions and exist over longer distance scales.  The length and complexity of this undertaking requires us to divide the calculation into multiple parts.  In this work (``Part I'') we will set up the formalism, construct the correlation functions and cross sections, and identify the wave function and Wilson line building blocks for these correlations.  By studying the long-distance asymptotics of these correlations, we will identify the mechanisms and associated length scales which characterize different physical regimes.  In subsequent publications, we will present the detailed set of Wilson line operators that describe the interactions in these regimes, solving for the correlation functions numerically and analytically.

The rest of this paper is organized as follows.  In Sec.~\ref{sec:formulation} we construct the definitions of the correlation functions and the coordinate-space cross-sections.  In Sec.~\ref{sec:ampl}, we construct the elementary $q \bar q$ production amplitude from the fundamental building blocks of light-front wave functions and Wilson line interactions, and we analyze how these building blocks are combined to form the single-pair production and double-pair production cross-sections.  In Sec.~\ref{sec:lengths} we analyze the long-distance asymptotics of various channels in the single- and double-pair production cross-sections to determine the length scales associated with the various types of correlations.  In Sec.~\ref{sec:concl} we present explicit results for $q \bar q$ correlations and baryon number correlations for single-pair production, summarize the overall physical picture of both single- and double-pair correlations, and conclude by outlining the relation of the analysis presented here in ``Part I'' to future work to follow.

%
\section{General Formulation}
\label{sec:formulation}
%

%
\subsection{Definitions: Quark, Antiquark, and Baryon Number Correlations}
\label{sec:defns} 
%

Let us construct the definitions of the quark/quark, quark/antiquark, antiquark/antiquark, and baryon number correlation functions following the treatment of \cite{Kittel:2005fu}.  Let $\Omega$ denote a kinematic window about mid-rapidity; the single- and double-inclusive probability densities $\rho_1^i$ and $\rho_2^{i j}$ to produce particles of species $i, j$ with kinematics $\omega_1 , \omega_2 \in \Omega$ are given by
\begin{align} \label{e:prob1}
  \rho_1^i (\omega_1) &= \frac{1}{\sigma_{inel}} \: \frac{d\sigma^i}{d \omega_1} 
  \notag \\
  \rho_2^{i j} (\omega_1 , \omega_2) &= \frac{1}{\sigma_{inel}} \: 
  \frac{d\sigma^{i j}}{d \omega_1 d \omega_2} .
\end{align}
Here $\sigma_{inel}$ is the total inelastic cross-section to produce any particles in the mid-rapidity window $\Omega$; at lowest order in perturbation theory, this corresponds to single-inclusive gluon production.  Integrating these probability densities over the phase space $\Omega$ yields the event-averaged number of each type of particle, 
\begin{align} 
\label{e:prob2}
  \int_{\Omega} d \omega_1 \, \rho_1^i (\omega_1) &= \langle n^i \rangle_{ev}
  \notag \\ 
  \int_{\Omega} d \omega_1 \, d \omega_2 \, \rho_2^{i j} (\omega_1, \omega_2) &= \langle n^i (n^j - \delta^{i j}) \rangle_{ev} ,
\end{align}
such that $\rho_1$ and $\rho_2$ simply correspond to multiplicities:
\begin{align} \label{e:prob3}
  \rho_1^i (\omega_1) &= \frac{1}{\sigma_{inel}} \frac{d\sigma^i}{d \omega_1} = 
  \left\langle \frac{d n^i}{d \omega_1} \right\rangle_{ev} 
  \notag \\
  \rho_2^{i j} (\omega_1 , \omega_2) &= \frac{1}{\sigma_{inel}} \frac{d\sigma^{i j}}{d \omega_1 \, d \omega_2} = 
  \left\langle \frac{d n^i}{d \omega_1} \frac{d n^j}{d \omega_2} \right\rangle_{ev} - \delta^{i j} \delta(\omega_1 - \omega_2) 
  \left\langle \frac{d n^i}{d \omega_1}  \right\rangle_{ev} ,
\end{align}
where $\langle \cdots \rangle_{ev}$ represents an average over events.

With the help of the generic formulas \eqref{e:prob3}, we can write down expressions for the average number of quark/quark ($q q$), quark/antiquark ($q \bar q$), and antiquark/antiquark ($\bar q \bar q$) pairs produced at mid-rapidity.  Throughout this paper we will denote the longitudinal degrees of freedom in terms of light-front components $v^\pm \equiv \frac{1}{\sqrt 2} (v^0 \pm v^3)$ and transverse degrees of freedom in terms of vectors $\bm{v} \equiv (v_\bot^1 , v_\bot^2)$ with magnitudes $v_T \equiv | \bm{v} |$.  The average number of particles (which may be either quarks or antiquarks) with longitudinal momenta $k_1^+ , k_2^+$ and transverse positions $\bm{B}_1 , \bm{B}_2$ are then given by
\begin{subequations} \label{e:prob4}
\begin{align} 
  \left\langle k_1^+ \frac{d n^q}{d^2 B_1 \, dk_1^+} \: 
  k_2^+ \frac{d n^q}{d^2 B_2 \, dk_2^+}  \right\rangle_{ev} &=
  \frac{1}{\sigma_{inel}} \Bigg[ k_1^+ k_2^+ \frac{d\sigma^{q q}}{d^2 B_1 \, dk_1^+ \: d^2 B_2 \, dk_2^+} 
  \notag \\ &
  + \delta^{(2)} (\bm{B}_1 - \bm{B}_2) \, k_2^+ \delta(k_1^+ - k_2^+) \:\:
  k_1^+ \frac{d\sigma^q}{d^2 B_1 \, dk_1^+} \Bigg]
  \\
  \left\langle k_1^+ \frac{d n^{q}}{d^2 B_1 \, dk_1^+} \: 
  k_2^+ \frac{d n^{\bar q}}{d^2 B_2 \, dk_2^+}  \right\rangle_{ev} &=
  \frac{1}{\sigma_{inel}} \Bigg[ k_1^+ k_2^+ \frac{d\sigma^{q \bar q}}
  {d^2 B_1 \, dk_1^+ \: d^2 B_2 \, dk_2^+} \Bigg]
  \\
  \left\langle k_1^+ \frac{d n^{\bar q}}{d^2 B_1 \, dk_1^+} \: 
  k_2^+ \frac{d n^{\bar q}}{d^2 B_2 \, dk_2^+}  \right\rangle_{ev} &=
  \frac{1}{\sigma_{inel}} \Bigg[ k_1^+ k_2^+ \frac{d\sigma^{\bar q \bar q}}
  {d^2 B_1 \, dk_1^+ \: d^2 B_2 \, dk_2^+} 
  \notag \\ &
  + \delta^{(2)} (\bm{B}_1 - \bm{B}_2) \, k_2^+ \delta(k_1^+ - k_2^+) \:\:
  k_1^+ \frac{d\sigma^{\bar q}}{d^2 B_1 \, dk_1^+} \Bigg]
\end{align}
\end{subequations}
In the same way, we can use the expressions \eqref{e:prob4} to construct the expectation values for conserved charges such as  baryon
number\footnote{The calculations presented here reflect the initial stages of hadronic collisions, before hadronization takes place.  As such, we refer more precisely here to ``parton-level baryon number,'' in which each quark / antiquark carries $\pm\frac{1}{3}$ units of conserved baryon number ``charge.''}
$\mathcal{B} \equiv \frac{1}{3} \sum_f ( n^{q_f} - n^{\bar{q}_f})$, with $\sum_f$ a sum over relevant quark flavors: 
\pagebreak
\begin{align} \label{e:prob5}
  \bigg\langle k_1^+ \frac{d \mathcal{B}}{d^2 B_1 dk_1^+} &\: 
  k_2^+ \frac{d \mathcal{B}}{d^2 B_2 dk_2^+} \bigg\rangle_{ev} 
  \notag \\ \notag \\ &
  \equiv \frac{1}{9} \sum_{f \, f'}
  \left\langle \left( k_1^+ \frac{d n^{q_f}}{d^2 B_1 dk_1} - k_1^+ \frac{d n^{\bar{q}_f}}{d^2 B_1 dk_1} \right) 
  \, \left( k_2^+ \frac{d n^{q_{f'}}}{d^2 B_2 dk_2^+} - k_2^+ \frac{d n^{\bar{q}_{f'}}}{d^2 B_2 dk_2^+} \right)
  \right\rangle_{ev}
  \notag \\ \notag \\ & =
  \frac{1}{9 \, \sigma_{inel}} \sum_{f \, f'} \bigg[ 
  k_1^+ k_2^+ \frac{d\sigma^{q_f q_{f'}}}{d^2 B_1  dk_1^+ \, d^2 B_2 dk_2^+} -
  k_1^+ k_2^+ \frac{d\sigma^{q_f \bar{q}_{f'}}}{d^2 B_1  dk_1^+ \, d^2 B_2 dk_2^+} 
  \notag \\ & 
  - k_1^+ k_2^+ \frac{d\sigma^{q_{f'} \bar{q}_f}}{d^2 B_2  dk_2^+ \, d^2 B_1 dk_1^+} 
  + k_1^+ k_2^+ \frac{d\sigma^{\bar{q}_{f} \bar{q}_{f'}}}{d^2 B_1  dk_1^+ \, d^2 B_2 dk_2^+} \bigg] 
  \notag \\ & 
  + \delta^{(2)} (\bm{B}_1 - \bm{B}_2) \, k_2^+ \delta(k_1^+ - k_2^+) \: \frac{1}{9 \, \sigma_{inel}} \sum_f 
  \bigg[ k_1^+ \frac{d\sigma^{q_f}}{d^2 B_1 dk_1^+} + k_1^+ \frac{d\sigma^{\bar{q}_f}}{d^2 B_1 dk_1^+} \bigg].
\end{align}

At different phase-space points $(\bm{B}_1 , k_1^+) \neq (\bm{B}_2 , k_2^+)$, the expectation values above describe nonlocal correlations in three dimensions, with only the double-inclusive cross-sections contributing. There is also an additional delta function contribution at $(\bm{B}_1 , k_1^+)  \rightarrow (\bm{B}_2 , k_2^+)$, with the weight of that delta function reflecting the strength of the local fluctuations (variance).  We define the associated correlation functions for these various quantities as
\begin{align} \label{e:corr1}
  \mathcal{C}_{i j} (\bm{B}_1 , k_1^+ \, ; \, \bm{B}_2 , k_2^+) & \equiv
  \left\langle k_1^+\frac{d n^i}{d^2 B_1 dk_1^+} \: k_2^+ \frac{d n^j}{d^2 B_2 dk_2^+} \right\rangle_{ev} 
  \notag \\ &
  - \left\langle k_1^+ \frac{d n^i}{d^2 B_1 dk_1^+} \right\rangle_{ev} 
  \left\langle k_2^+ \frac{d n^j}{d^2 B_2 dk_2^+} \right\rangle_{ev} ,
\end{align}
with $n^i , n^j \in \{ n^q , n^{\bar q} , \mathcal{B} \}$, where we assume that $(\bm{B}_1 , k_1^+) \neq (\bm{B}_2 , k_2^+)$ to exclude the delta function term.  The first term in \eqref{e:corr1} describes the correlated production of two particles, from which the uncorrelated background in the second term is subtracted.  While this definition of the correlation function is perhaps the simplest, it is by no means unique.  Some references \cite{Kovchegov:2012nd, Khachatryan:2010gv}, for example, rescale the correlation function to make it dimensionless, and others \cite{Plumberg:2013nga} do not subtract the uncorrelated background.  We will explore the features of different definitions of the correlation function in future work, but for now, the simple definition \eqref{e:corr1} will suffice.

The production of net baryon number at mid-rapidity is suppressed~\cite{Itakura:2003jp, Kharzeev:1996sq, Capella:1996th, Rossi:1977cy, Altinoluk:2016vax} by the center-of-mass energy squared $s$ 
\begin{align} \label{e:netB}
  k_1^+ \left\langle \frac{d \mathcal{B}}{d^2 B_1 d k_1^+} \right\rangle_{ev} \sim \ord{\frac{1}{s}} ,
\end{align}
corresponding to baryon stopping, in which valence quarks lose nearly all of their energy in the collision and are produced at mid-rapidity.  Instead, it is far more preferable at high energies for these valence degrees of freedom in the colliding nuclei to punch right through each other and continue down the beam pipe, with the mid-rapidity region instead being populated by soft gluon bremsstrahlung.  With eikonal accuracy, then, the inclusive probabilities to produce a quark or antiquark at mid-rapidity are equal and given at leading order by integrating out the spectator in $q \bar q$ pair production:
\begin{align} \label{e:uncorr}
\left\langle k_1^+ \frac{d n^q}{d^2 B_1 dk_1^+} \right\rangle_{ev} &= 
\left\langle k_1^+ \frac{d n^{\bar q}}{d^2 B_1 dk_1^+} \right\rangle_{ev}  =
\frac{1}{\sigma_{inel}} \: k_1^+ \frac{d\sigma^q}{d^2 B_1 dk_1^+} 
\notag \\ & \overset{L.O.}{=}
\int\frac{dk_2^+}{k_2^+} \int d^2 B_2 \, \left( \frac{1}{\sigma_{inel}} k_1^+ k_2^+ \frac{d\sigma^{q \bar q}}{d^2 B_1 dk_1^+ \, d^2 B_2 dk_2^+} \right).
\end{align}
This single-inclusive cross-section defines the uncorrelated baseline for the correlation functions.  The correlation functions of interest are therefore given by
\begin{subequations} \label{e:corrs}
 \begin{align} \label{e:qcorr1}
  \mathcal{C}_{q q} (\bm{B}_1, k_1^+ \, ; \, \bm{B}_2, k_2^+) &= 
  \frac{1}{\sigma_{inel}} \: k_1^+ k_2^+ \, \frac{d\sigma^{q q}}{d^2 B_1 dk_1^+ \, d^2 B_2 dk_2^+}
   \notag \\ &
   - \left( \frac{1}{\sigma_{inel}} \: k_1^+ \frac{d\sigma^q}{d^2 B_1 dk_1^+} \right)
  \left( \frac{1}{\sigma_{inel}} \: k_2^+ \frac{d\sigma^q}{d^2 B_2 dk_2^+} \right)
   \\ \notag \\ \label{e:qaqcorr1} 
   \mathcal{C}_{q \bar q} (\bm{B}_1, k_1^+ \, ; \, \bm{B}_2, k_2^+) &= 
   \frac{1}{\sigma_{inel}} \: k_1^+ k_2^+ \, \frac{d\sigma^{q \bar q}}{d^2 B_1 dk_1^+ \, d^2 B_2 dk_2^+}
   \notag \\ &
  - \left( \frac{1}{\sigma_{inel}} \: k_1^+ \frac{d\sigma^q}{d^2 B_1 dk_1^+} \right)
  \left( \frac{1}{\sigma_{inel}} \: k_2^+ \frac{d\sigma^q}{d^2 B_2 dk_2^+} \right)
   \\ \notag \\ \label{e:aqcorr1}
   \mathcal{C}_{\bar q \bar q} (\bm{B}_1, k_1^+ \, ; \, \bm{B}_2, k_2^+) &= 
   \frac{1}{\sigma_{inel}} \: k_1^+ k_2^+ \, \frac{d\sigma^{\bar q \bar q}}{d^2 B_1 dk_1^+ \, d^2 B_2 dk_2^+}
   \notag \\ &
   - \left( \frac{1}{\sigma_{inel}} \: k_1^+ \frac{d\sigma^q}{d^2 B_1 dk_1^+} \right)
  \left( \frac{1}{\sigma_{inel}} \: k_2^+ \frac{d\sigma^q}{d^2 B_2 dk_2^+} \right)
   \\ \notag \\ \label{e:Bcorr1}
   \mathcal{C}_{\mathcal{B} \mathcal{B}} (\bm{B}_1, k_1^+ \, ; \, \bm{B}_2, k_2^+) &= \frac{1}{9} \sum_{f \, f'}
   \Bigg( \mathcal{C}_{q_f q_{f'}} (\bm{B}_1, k_1^+ \, ; \, \bm{B}_2, k_2^+) - 
   \mathcal{C}_{q_f \bar{q}_{f'}} (\bm{B}_1, k_1^+ \, ; \, \bm{B}_2, k_2^+) 
   \notag \\ & \hspace{1cm}
   - \mathcal{C}_{q_{f'} \bar{q}_f} (\bm{B}_2, k_2^+ \, ; \, \bm{B}_1, k_1^+)
   + \mathcal{C}_{\bar{q}_f \bar{q}_{f'}} (\bm{B}_1, k_1^+ \, ; \, \bm{B}_2, k_2^+) \Bigg)
 \end{align}
\end{subequations}
for $(\bm{B}_1 , k_1^+) \neq (\bm{B}_2 , k_2^+)$.

%
\subsection{Structure of the Cross Sections}
%

%
\begin{figure}
\includegraphics[width=0.8\textwidth]{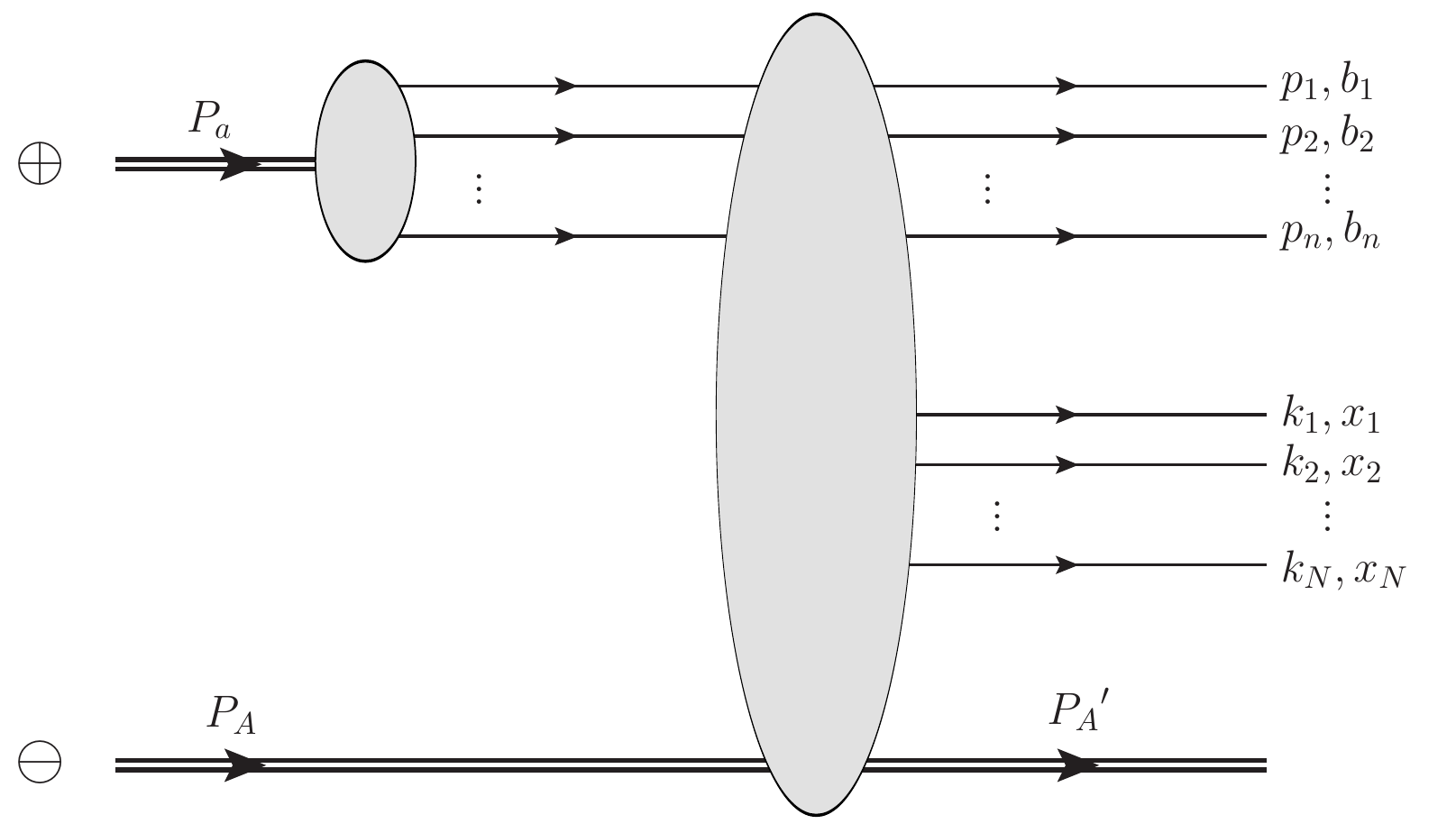}
\caption{Illustration of the high-energy scattering process.  The valence degrees of freedom from the light projectile ion are produced in the forward direction, the remnants from the heavy target ion are produced in the backward direction, and a number of soft particles are produced at mid-rapidity. The momenta and coordinates of the valence quarks are denoted by $p_i$ and $b_i$ respectively. The momenta and coordinates of the particles produced at mid-rapidity are denoted by $k_i$ and $x_i$ respectively. } 
\label{f:Particles}
\end{figure}
%

Next we will explicitly construct the two-particle-inclusive cross-sections which enter the correlation functions \eqref{e:corrs}.  These cross-sections which define the spatial correlations are somewhat unusual in that they are differential in the transverse positions of the produced (anti)quarks, rather than their momenta, so it is useful to construct them directly from textbook principles (e.g.~\cite{Peskin:1995ev}).  

Consider the collision shown in Fig.~\ref{f:Particles} between a light ``projectile'' nucleus with $a$ nucleons moving with large momentum $P_a^+$ along the light-cone $\oplus$ direction and a heavy ``target'' nucleus with $A$ nucleons moving with large momentum $P_A^-$ along the light-cone $\ominus$ direction.  The paradigm of ``heavy-light ion collisions'' corresponds to the regime in which the target nucleus is so heavy that its density-enhanced corrections must be resummed to all orders, $\alpha_s^2 A^{1/3} \sim \ord{1}$, while the density-enhanced corrections from the projectile nucleus are included order by order in perturbation theory.  More precisely, the region of interest is $\alpha_s \ll \alpha_s^2 a^{1/3} \ll 1$, such that corrections enhanced by the density $\sim a^{1/3}$ of the light ion are more important than genuine quantum loops, but not so large that they must be simultaneously resummed.  In the inclusive final state, we will identify a number of particles traveling in three distinct regions: the ``forward'' fragmentation region of the light nucleus $a$, the ``backward'' fragmentation region of the heavy nucleus $A$, and the mid-rapidity region between the two.  In the forward region, we consider a set of $n$ active valence quarks in the light nucleus $a$ with momenta
$\{ p_j \}_{j = 1}^{n}$.\footnote{For a nucleus with many nucleons, it is combinatorically preferable for each valence quark to come from a different nucleon, so these valence quarks act as proxies for nucleons in the light nucleus.}  
For our purposes, we will only need up to $n = 2$ independent valence quarks from the light nucleus.  In the mid-rapidity region, we consider a set of $N$ particles with momenta $\{ k_i \}_{i = 1}^N$ which will generate the correlations of interest, and in the backward region we consider a generic remnant state with momentum $P_A^\prime$.  The standard momentum-space expression for the cross section is then
\begin{align}
d\sigma_{aA} &= \frac{1}{2 s} \left( \prod_{i = 1}^N \frac{d^2 k_i \, dk_i^+}{2(2\pi)^3 k_i^+} \right) 
\left( \prod_{j = 1}^n \frac{d^2 p_j \, dp_j^+}{2(2\pi)^3 p_j^+} \right) 
\frac{d^2 P_A^\prime \, dP_A^{\prime \, -}}{2(2\pi)^3 P_A^{\prime -}} 
\notag \\ & \hspace{0.5cm} \times
\: |\mathcal{M}_{aA}|^2 \: (2\pi)^4 \delta^4 \left( P_a + P_A - P_A^\prime - \sum_{i=1}^N k_i 
- \sum_{j=1}^n p_j \right).
\end{align}

With eikonal accuracy, the particles traveling in the forward direction have negligible minus momenta, $P_a^- , p_j^- \sim 0$, particles traveling in the backward direction have negligible plus momenta, $P_A^+ , P_A^{\prime \, +} \sim 0$, and particles at mid-rapidity have both negligible plus and minus momenta: $k_i^+ , k_i^- \sim 0$.  This allows us to approximate the momentum-conserving delta function by the product of $\delta(P_a^+ - \sum_{j=1}^n p_j^{+})$, $\delta(P_A^- - P_A^{\prime -})$, and $\delta^2 (\bm{P}_A^\prime + \sum_{i=1}^N \bm{k}_i + \sum_{j=1}^n \bm{p}_j)$ and performing the $d^2 P_A^\prime \, dP_A^{\prime \, -}$ integrals directly, we obtain
\begin{align}
d\sigma_{aA} &= \left( \prod_{i = 1}^N \frac{d^{2+} k_i}{2(2\pi)^3 k_i^+} \right) \int
\left( \prod_{j = 1}^n \frac{d^2 p_j}{(2\pi)^2}  \frac{dz_j}{4\pi \, z_j}\right) \:
4\pi \, \delta (1 - \sum_{j = 1}^n z_j) \: | \mathcal{A}_{aA} |^2 ,
\end{align}
where $\mathcal{A}_{aA} \equiv \mathcal{M}_{aA} / 2 s$ with $s = 2 P_a^+ P_A^-$ and $z_j = p_j^{ +} / P_a^+$.  Next we Fourier transform the amplitude to transverse coordinate space in the produced particles; that is, we insert
\begin{align}
\mathcal{A}_{aA} ( \{ \bm{p}_j , z_j \}_{j=1}^n ,\{ \bm{k}_i , k_i^+ \}_{i=1}^N) &=
\int \left( \prod_{i = 1}^N d^2 x_i \: e^{-i \bm{k}_i \cdot \bm{x}_i} \right)
\int \left( \prod_{j = 1}^n d^2 b_i \: e^{-i \bm{p}_i \cdot \bm{b}_i} \right)
\notag \\ & \hspace{0.5cm} \times
\tilde{\mathcal{A}}_{aA} ( \{ \bm{b}_j , z_j \}_{j=1}^n ,\{ \bm{x}_i , k_i^+ \}_{i=1}^N)
\end{align}
and perform the transverse momentum integrals to obtain
\begin{align}
d\sigma_{aA} &= \left( \prod_{i = 1}^N d^2 x_i \frac{dk_i^+}{4\pi k_i^+} \right) \int
\left( \prod_{j = 1}^n d^2 b_j \frac{dz_j}{4\pi \, z_j}\right) \:
4\pi \, \delta (1 - \sum_{j = 1}^n z_j) \: | \tilde{\mathcal{A}}_{aA} |^2 .
\end{align}
Note that, having integrated out the transverse momenta, the coordinates of the produced particles are the same in the amplitude and complex-conjugate amplitude; this simplification makes the problem of multi-particle correlations much more tractable in coordinate space than in momentum space.

In the eikonal approximation, neither the transverse coordinates $\{ \bm{b}_j \}$ nor the longitudinal momenta $\{z_j P_a^+\}$ of the light-nucleus valence quarks are changed by the interaction with the target; they therefore correspond to the many-body wave function $\Psi_a$ of valence quarks in the light nucleus:
\begin{align}
\tilde{\mathcal{A}}_{aA} ( \{ \bm{b}_j , z_j \}_{j=1}^n ,\{ \bm{x}_i , k_i^+ \}_{i=1}^N) &=
\Psi_a ( \{ \bm{b}_j , z_j \}_{j=1}^n ) \:
\tilde{\mathcal{A}}_{\{N\} A} ( \{ \bm{b}_j \}_{j=1}^n ,\{ \bm{x}_i , k_i^+ \}_{i=1}^N),
\end{align}
where $\tilde{\mathcal{A}}_{\{N\} A} ( \{ \bm{b}_j , z_j \}_{j=1}^n ,\{ \bm{x}_i , k_i^+ \}_{i=1}^N)$ is the scaled amplitude for a set $\{ N \}$ of valence quarks (or nucleons) to scatter on the target nucleus.  Note that, in the quasi-classical approximation (where the collision energy is not so large that quantum evolution need be considered), the amplitude $\tilde{\mathcal{A}}_{\{N\} A}$ is independent of the momentum fractions $\{ z_j \}$.  We can also insert a Fourier transformation of the cross-section to impact parameter space $\int d^2 B \, e^{+ i \bm{P}_a \cdot \bm{B}}$, with $\bm{P}_a = 0$ in the center-of-mass frame.  Using this, we have
\begin{align} 
d\sigma_{aA} &= \left( \prod_{i = 1}^N d^2 x_i \frac{dk_i^+}{4\pi k_i^+} \right) 
\int d^2 B \left( \prod_{j = 1}^n d^2 b_j \right) \:
\notag \\ &
\times \left[ \int \left( \prod_{j = 1}^n \frac{dz_j}{4\pi \, z_j} \right) 4\pi \, \delta (1 - \sum_{j = 1}^n z_j) \: 
\left| \Psi_a ( \{ \bm{b}_j - \bm{B}, z_j \}_{j=1}^n ) \right|^2 \right]
| \tilde{\mathcal{A}}_{ \{ N \}A} |^2 ,
\end{align}
and by integrating the $z_j$ dependence out of the wave functions, we can replace the wave functions of the light nucleus in terms of the nuclear density profile $T_a (\bm{b})$:
\begin{align} \label{e:genxsec}
d\sigma_{aA} &= \left( \prod_{i = 1}^N d^2 x_i \frac{dk_i^+}{4\pi k_i^+} \right) 
\int d^2 B \left( \prod_{j = 1}^n d^2 b_j \, T_a(\bm{b}_j - \bm{B}) \right) \: | \tilde{\mathcal{A}}_{ \{ N \}A} |^2 .
\end{align}
The nuclear density profiles are normalized such that $\int d^2 b \, T_a (\bm{b}) = a$, and since the $d^2 b$ integral in this normalization is proportional to the transverse area of the light nucleus, this implies that $T_a (\bm{b}) \sim a^{1/3}$. Each explicit factor of $T_a$ above therefore reflects the combinatoric enhancement by $a^{1/3}$ associated with independent nucleons participating in the scattering process.

With the general form of the coordinate-space cross section \eqref{e:genxsec}, we can construct the two-particle-inclusive cross-sections entering the correlations \eqref{e:corrs}.  There are two fundamental cases of interest: single-pair production, in which one active nucleon $(n=1)$ from the light projectile nucleus radiates a soft $q \bar q$ pair $(N=2)$; and double-pair production, in which two active nucleons $(n=2)$ from the light projectile nucleus independently radiate two $q \bar q$ pairs $(N=4)$.  These different processes contribute to different correlations, and the dominant process may change as a function of distance.  Single-pair production, for instance, contributes to $\mathcal{C}_{q \bar q}$ but not $\mathcal{C}_{q q}$, while double-pair production contributes to all correlations but is sub-leading compared to single-pair production where applicable.  

\begin{figure}
\includegraphics[width=0.5\textwidth]{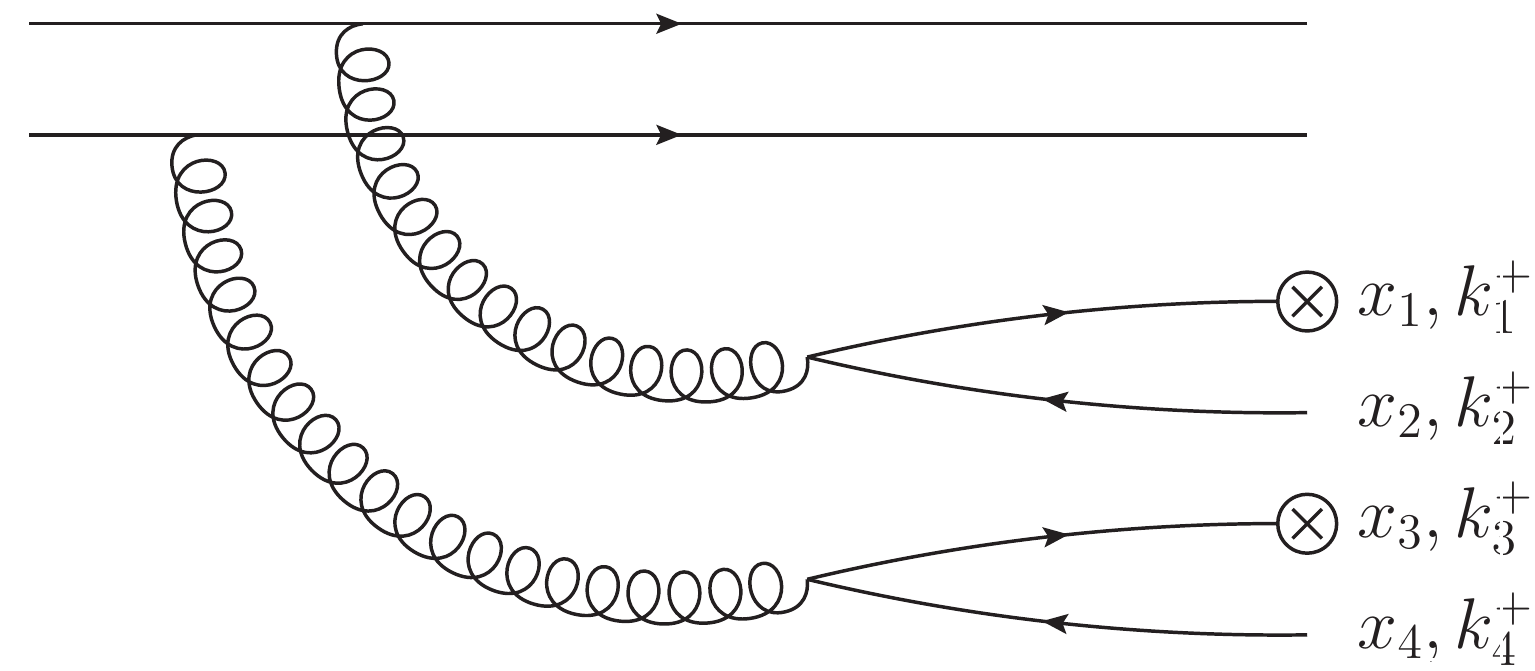}
\caption{Coordinate labels for double-pair production.  Quarks are produced at $\bm{x}_1 , k_1^+$ and $\bm{x}_3 , k_3^+$ and antiquarks are produced at $\bm{x}_2 , k_2^+$ and $\bm{x}_4 , k_4^+$.  The crossed vertices represent the tagging delta functions \eqref{e:Ztags} which assign two of the particles to the external coordinates $\bm{B}_1 , K_1^+$ and $\bm{B}_2 , K_2^+$.  The tagged combination shown here corresponds to $Z^{(q q)}$, for example.  Note that the pairs to which the (anti)quarks belong can become entangled in the complex-conjugate amplitude.} 
\label{f:Dbl_Pair_Coords}
\end{figure}

For single-pair production, which contributes only to $q \bar q$ correlations, we can straightforwardly write
\begin{align} \label{e:CSsingle}
k_1^+ k_2^+ \, \frac{d\sigma^{q \bar q}_{single}}{d^2 B_1 dk_1^+ \, d^2 B_2 dk_2^+} &= 
\frac{1}{(4\pi)^2} \int d^2 B \, d^2 b_1 \, T_a (\bm{b_1} - \bm{B}) \: 
| \tilde{\mathcal{A}}_{N A} (\bm{b}_1 ; \bm{B}_1 , k_1^+ , \bm{B}_2 , k_2^+) |^2
\notag \\ &=
\frac{a}{(4\pi)^2} \int d^2 b_1 \: 
| \tilde{\mathcal{A}}_{N A} (\bm{b}_1 ; \bm{B}_1 , k_1^+ , \bm{B}_2 , k_2^+) |^2 ,
\end{align}
where we have performed the integral over the impact parameter $\bm{B}$ to generate a factor of $a$.  For double pair production, it is convenient to formulate the cross-sections by integrating over all the (anti)quarks in the final state, assigning the tagged particles through the use of delta functions as illustrated in Fig.~\ref{f:Dbl_Pair_Coords}:
\begin{subequations} \label{e:Ztags}
\begin{align}
Z^{(q \bar q)} &\equiv \frac{1}{4} K_1^+ K_2^+ \left[  \delta^2 (\bm{x}_1 - \bm{B}_1) \, \delta(k_1^+ - K_1^+) +
\delta^2 (\bm{x}_3 - \bm{B}_1) \, \delta(k_3^+ - K_1^+) \right] 
\notag \\ & \hspace{2cm} \times
\left[  \delta^2 (\bm{x}_2 - \bm{B}_2) \, \delta(k_2^+ - K_2^+) +
\delta^2 (\bm{x}_4 - \bm{B}_2) \, \delta(k_4^+ - K_2^+) \right]
\\
Z^{(q q)} &\equiv \frac{1}{2} K_1^+ K_2^+ \left[  \delta^2 (\bm{x}_1 - \bm{B}_1) \, \delta(k_1^+ - K_1^+) \:
\delta^2 (\bm{x}_3 - \bm{B}_2) \, \delta(k_3^+ - K_2^+) \right. 
\notag \\ & \hspace{2cm} +
\left.  \delta^2 (\bm{x}_3 - \bm{B}_1) \, \delta(k_3^+ - K_1^+) \:
\delta^2 (\bm{x}_1 - \bm{B}_2) \, \delta(k_1^+ - K_2^+) \right]
\\
Z^{(\bar q \bar q)} &\equiv \frac{1}{2} K_1^+ K_2^+ \left[  \delta^2 (\bm{x}_2 - \bm{B}_1) \, \delta(k_2^+ - K_1^+) \:
\delta^2 (\bm{x}_4 - \bm{B}_2) \, \delta(k_4^+ - K_2^+) \right. 
\notag \\ & \hspace{2cm} +
\left.  \delta^2 (\bm{x}_4 - \bm{B}_1) \, \delta(k_4^+ - K_1^+) \:
\delta^2 (\bm{x}_2 - \bm{B}_2) \, \delta(k_2^+ - K_2^+) \right].
\end{align}
\end{subequations}
%
%
%
%
%
Note the inclusion of the symmetry factor for identical tagged particles.  Then the cross-section for double-pair production can be written in a standard form for all three observables as
\begin{align} \label{e:CSdouble}
K_1^+ K_2^+ \frac{d\sigma_{double}}{d^2 B_1 dK_1^+ \, d^2 B_2 dK_2^+} &= 
\int d^2 B \, d^2 b_1 \, d^2 b_2 \, T_a (\bm{b}_1 - \bm{B}) \, T_a (\bm{b}_2 - \bm{B}) \:
\int \left( \prod_{i = 1}^4 d^2 x_i \frac{dk_i^+}{4\pi k_i^+} \right)
\notag \\ & \hspace{-1cm} \times
| \tilde{\mathcal{A}}_{N N A} (\bm{b}_1 , \bm{b}_2 ; \{ \bm{x}_i , k_i^+ \}_{i=1}^4) |^2 
Z(\bm{B}_1 , K_1^+ , \bm{B}_2 , K_2^+ , \{ \bm{x}_i , k_i^+ \}_{i=1}^4) .
\end{align}
Comparing the single- and double-pair cross-sections \eqref{e:CSsingle} and \eqref{e:CSdouble}, we see the relative power-counting of these two processes,
\begin{subequations}
\begin{align}
d\sigma_{single} &\sim T_a (\bm{b}_1 - \bm{B}) \left| \tilde{\mathcal{A}}_{N A} \right|^2 
\sim \alpha_s^2 a^{1/3}
\\ 
d\sigma_{double} &\sim T_a (\bm{b}_1 - \bm{B}) \, T_a (\bm{b}_2 - \bm{B}) 
\left| \tilde{\mathcal{A}}_{N N A} \right|^2 
\sim (\alpha_s^2 a^{1/3})^2 ,
\end{align}
\end{subequations}
where we recall that $T_a (\bm{b}) \sim a^{1/3}$ and the amplitude for each radiated pair brings in a factor of $\alpha_s$.  These considerations imply that, in the heavy-light regime, double-pair production is suppressed relative to single-pair production by a factor of $\alpha_s^2 a^{1/3} \ll 1$.  Note also that, for double-pair production, the $d^2 B$ integral is no longer trivial and itself can induce geometric correlations between the pairs \cite{Kovchegov:2012nd}.  Finally, to convert the cross sections \eqref{e:CSsingle} and \eqref{e:CSdouble} into the correlation functions \eqref{e:corrs}, we need to normalize by the total $a A$ inelastic cross-section $\sigma_{inel}$.  At lowest order in perturbation theory, $\sigma_{inel}$ corresponds to the total cross-section for single-inclusive gluon production at mid-rapidity, which is a standard textbook result \cite{Kovchegov:2012mbw}.

%
\section{Construction of the Amplitude}
\label{sec:ampl}
%

The core of the calculation is the construction of the scattering amplitudes $\tilde{\mathcal{A}}_{\{N\} A}$ for the production of one or two $q \bar q$ pairs.  We will perform this calculation using the formalism of light-front perturbation theory (LFPT) \cite{Bjorken:1970ah, Lepage:1980fj, Brodsky:1997de} and choose the $A^+ = 0$ light-cone gauge, in which the $q \bar q$ pairs are radiated from the light projectile nucleus $a$.  In this framework, the amplitude $\tilde{\mathcal{A}}_{\{N\} A}$ is given by a product of light-front wave functions $\Psi$ describing the emission of the $q \bar q$ pair and Wilson lines describing $SU(N_c)$ color rotations (with $N_c$ the number of colors) in the fundamental or adjoint representations.  Here we will construct these building blocks of the amplitude and contract them in all possible ways to obtain the cross-sections \eqref{e:CSsingle} and \eqref{e:CSdouble}.

%
\subsection{Building Blocks: Wave Functions and Wilson Lines}
%

%
\begin{figure}
\includegraphics[width=\textwidth]{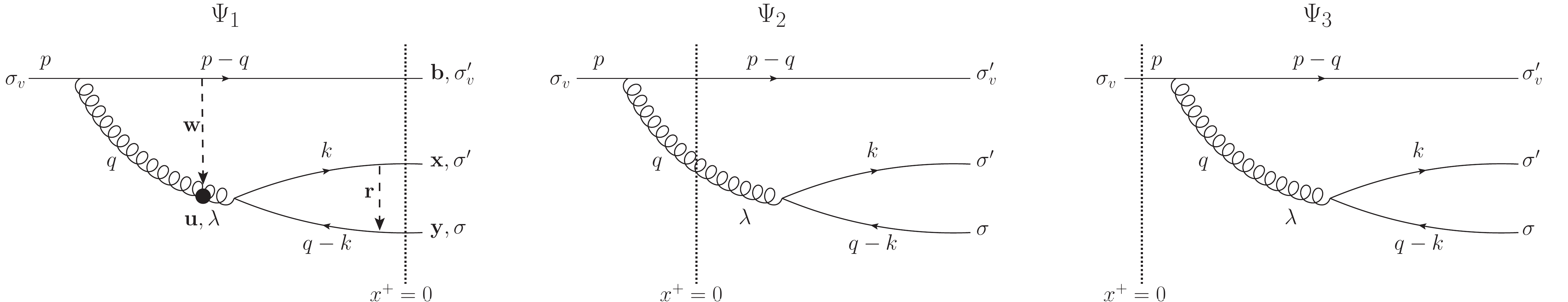}
\caption{ \label{f:WF}
The light-front wave functions for the radiation of a $q \bar q$ pair, depending on the placement of light-front time $t_{LF} = x^+ = 0$.  $\Psi_1$ describes the $q \bar q$ emission before $x^+ = 0$, $\Psi_3$ describes the radiation after $x^+ = 0$, and $\Psi_2$ describes the gluon emission before $x^+ = 0$ and the $q \bar q$ pair production after.  The transverse positions for the Fourier transform are also shown for $\Psi_1$.} 
\end{figure}
%

In a time-ordered formalism like LFPT, there are three distinct time orderings and three distinct wave functions for the production of a single $q \bar q$ pair, as shown in Fig.~\ref{f:WF}.  The wave functions in this language were previously calculated in \cite{Kovchegov:2006qn}, and we rederive them in Appendix~\ref{app:WFs}.  The results for the momentum-space wave functions are
\begin{subequations} \label{e:WFmom}
\begin{align}
\Psi_1 &= - 2 g^2 \frac{\sqrt{\alpha (1-\alpha)}}{( \bm{k} - \alpha \bm{q})_T^2 + m^2 + \alpha(1-\alpha) q_T^2}
\notag \\ & \hspace{1cm} \times
\bigg\{ \delta_{\sigma \, , \, -\sigma'} \left[ \alpha + (1-2\alpha) \frac{\bm{q} \cdot \bm{k}}{q_T^2} 
- i \sigma' \, \frac{\bm{q} \times \bm{k}}{q_T^2} \right] 
-m \sigma' \, \delta_{\sigma \sigma'} \left[ \frac{\bm{q}_\bot^1}{q_T^2} 
- i \sigma' \, \frac{\bm{q}_\bot^2}{q_T^2} \right] \bigg\}
\\ \notag \\ 
\Psi_2 &= -2 g^2 \frac{\sqrt{\alpha (1-\alpha)}}{( \bm{k} - \alpha \bm{q})_T^2 + m^2}
\bigg\{ \delta_{\sigma \, , \, -\sigma'} \left[ - (1-2\alpha) \frac{\bm{q} \cdot (\bm{k} - \alpha \bm{q})}{q_T^2} 
+ i \sigma' \, \frac{\bm{q} \times (\bm{k} - \alpha \bm{q})}{q_T^2} \right] 
\notag \\ & \hspace{4cm} +
m \sigma' \, \delta_{\sigma \sigma'} \left[ \frac{\bm{q}_\bot^1}{q_T^2} 
- i \sigma' \, \frac{\bm{q}_\bot^2}{q_T^2} \right] \bigg\}
\\ \notag \\ 
\Psi_3 &= - \Psi_1 - \Psi_2 ,
\end{align}
\end{subequations}
with the momenta labeled as in Fig.~\ref{f:WF} and $\alpha \equiv \frac{k_1^+}{q^+}$ the fraction of the pair momentum carried by the quark.  Here we explicitly keep the quark mass to consider the possibility of producing heavy quarks, since this mass characterizes the typical size of the $q \bar q$ separation.  We then Fourier transform into coordinate space using
\begin{align}
\tilde{\Psi}_i &\equiv \int\frac{d^2 k}{(2\pi)^2} \frac{d^2 (q-k)}{(2\pi)^2} \, e^{i \bm{k} \cdot (\bm{x} - \bm{b})} \,
e^{i (\bm{q} - \bm{k}) \cdot (\bm{y} - \bm{b})} \, \Psi_i (\alpha , \bm{q} , \bm{k} - \alpha \bm{q})
\notag \\ &=
\int\frac{d^2 (k-\alpha q)}{(2\pi)^2} \frac{d^2 q}{(2\pi)^2} \, e^{i (\bm{k} - \alpha \bm{q}) \cdot \bm{r}} \,
e^{i \bm{q} \cdot \bm{w}} \, \Psi_i (\alpha , \bm{q} , \bm{k} - \alpha \bm{q}),
\end{align}
where $\bm{r} \equiv \bm{x} - \bm{y}$ and $\bm{w} \equiv \bm{u} - \bm{b}$ with $\bm{u} \equiv \alpha \bm{x} + (1-\alpha) \bm {y}$ the position of the gluon at the pair center of momentum, as shown in Fig.~\ref{f:WF}.  This choice of variables makes use of the fact that the wave functions depend only on the center-of-mass momentum $\bm{q}$ of the pair and the relative momentum $(\bm{k} - \alpha \bm{q})$ of the pair when boosted to its rest frame.  Performing the Fourier transform we obtain the coordinate space expressions
\begin{subequations} \label{e:WFs}
\begin{align}
\tilde{\Psi}_1 &= \frac{g^2}{2 \pi^2} \sqrt{\alpha(1-\alpha)} \bigg\{
\delta_{\sigma \, , \, -\sigma'} \bigg[ F_2 (w_T, r_T, \alpha) \left( (1-2\alpha) \frac{\bm{w} \cdot \bm{r}}{w_T r_T}
- i \sigma' \, \frac{\bm{w} \times \bm{r}}{w_T r_T} \right) 
\notag \\ & \hspace{5cm} -
2 \alpha (1-\alpha) F_0 (w_T, r_T, \alpha) \bigg]
\notag \\ & \hspace{3cm} +
i m \sigma' \, \delta_{\sigma \sigma'} \, F_1 (w_T, r_T, \alpha) \left[ \frac{\bm{w}_\bot^1}{w_T} - 
i \sigma' \, \frac{\bm{w}_\bot^2}{w_T} \right] \bigg\}
\\ \notag \\
\tilde{\Psi}_2 &= -\frac{g^2}{2\pi^2} \sqrt{\alpha (1-\alpha)} \bigg\{
\delta_{\sigma \, , \, -\sigma'} \, \frac{m}{w_T} K_1 (m r_T) \left[ (1-2\alpha) \frac{\bm{w} \cdot \bm{r}}{w_T r_T}
- i \sigma' \, \frac{\bm{w} \times \bm{r}}{w_T r_T} \right]
\notag \\ & \hspace{3cm} +
i \sigma' \, \delta_{\sigma \sigma'} \, \frac{m}{w_T} K_0 (m r_T) \left[ \frac{\bm{w}_\bot^1}{w_T} - 
i \sigma' \, \frac{\bm{w}_\bot^2}{w_T} \right] \bigg\}
\\ \notag \\ 
\tilde{\Psi}_3 &= - \tilde{\Psi}_1 - \tilde{\Psi}_2 ,
\end{align}
\end{subequations}
where the functions $F_i (w_T, r_T, \alpha)$ are defined as in \cite{Kovchegov:2006qn},
\begin{subequations} \label{e:Fdefs}
\begin{align}
F_0 (w_T, r_T, \alpha) &\equiv \int\limits_0^\infty dq_T \, q_T \, J_0 (q_T w_T) \,
  K_0 \! \left( r_T \sqrt{m^2 + \alpha (1-\alpha) q_T^2} \right) \\
F_1 (w_T, r_T, \alpha) &\equiv \int\limits_0^\infty dq_T \, J_1 (q_T w_T) \,
  K_0 \! \left( r_T \sqrt{m^2 + \alpha (1-\alpha) q_T^2} \right) \\
F_2 (w_T, r_T, \alpha) &\equiv \int\limits_0^\infty dq_T \, J_1 (q_T w_T) \,
  K_1 \! \left( r_T \sqrt{m^2 + \alpha (1-\alpha) q_T^2} \right) \, \sqrt{m_q^2 + \alpha (1-\alpha) q_T^2}.
\end{align}
\end{subequations}
Note that $F_0$ and $F_2$ have dimensions of $M^2$, while $F_1$ has dimensions of $M^1$.  

We further note that the spin structure of the wave functions can be conveniently decomposed \cite{Kovchegov:2015zha} in terms of the Pauli matrices $[ \, \vec{\tau} \, ]$ as
\begin{align} \label{e:WFdecomp}
  \left[ \tilde{\Psi}_i (\bm w, \bm r , \alpha) \right]_{\sigma', - \sigma} &\equiv 
  \left[ \mathds 1 \right]_{\sigma', - \sigma} \mathcal{U}_i (\bm w, \bm r , \alpha) +
  \left[ \tau_3 \right]_{\sigma', - \sigma} \mathcal{L}_i  (\bm w, \bm r , \alpha)
  \notag \\ &
  + \left[ \bm{\tau} \right]_{\sigma', - \sigma} \times \bm{\mathcal{T}}_i   (\bm w, \bm r , \alpha) ,
\end{align}
with $\mathcal{U}$ , $\mathcal{L}$ , and $\bm{\mathcal{T}}$ denoting unpolarized, longitudinally polarized, and transversely polarized quarks, respectively.  Note that, in order to get a Pauli representation which transforms as a vector under 2D rotations, it is necessary to represent the antiquark quantum number as $(-\sigma)$, similar to the idea of an antiquark helicity used in \cite{Kovchegov:2006qn}.  This construction yields a transparent physical interpretation of the spin state of the $q \bar q$ pair, is independent of the spinor basis, and is explicitly invariant under rotations in the transverse plane \cite{Kovchegov:2015zha}.  The corresponding spatial wave functions are given by
\begin{subequations} \label{e:WFlabels}
\begin{align}
\mathcal{U}_1 (\bm w , \bm r, \alpha) &= \frac{g^2}{2\pi^2} \sqrt{\alpha(1-\alpha)} \Big[
  (1 - 2\alpha) \frac{\bm w \cdot \bm r}{w_T \, r_T} F_2 (w_T , r_T, \alpha) 
\notag \\ &-
2 \alpha (1-\alpha) F_0 (w_T , r_T, \alpha) \Big] 
\\
 \mathcal{L}_1 (\bm w , \bm r, \alpha) &= \frac{g^2}{2\pi^2} \sqrt{\alpha(1-\alpha)} \left[
- i \, \frac{\bm w \times \bm r}{w_T \, r_T} F_2 (w_T , r_T, \alpha) \right]
\\
 \bm{\mathcal{T}}_1 (\bm w , \bm r, \alpha) &= \frac{g^2}{2\pi^2} \sqrt{\alpha(1-\alpha)} \left[ 
\left( \frac{m}{w_T} F_1 (w_T, r_T, \alpha) \right) \bm w \right]
\\ \notag \\
\mathcal{U}_2 (\bm w , \bm r, \alpha) &= \frac{g^2}{2\pi^2} \sqrt{\alpha(1-\alpha)} \left[ 
-(1 - 2\alpha) \frac{\bm w \cdot \bm r}{w_T \, r_T} \, \frac{m}{w_T} K_1 (m r_T) \right]
\\
\mathcal{L}_2 (\bm w , \bm r, \alpha) &= \frac{g^2}{2\pi^2} \sqrt{\alpha(1-\alpha)} \left[  
i \, \frac{\bm w \times \bm r}{w_T \, r_T} \, \frac{m}{w_T} K_1 (m r_T) \right]
\\
\bm{\mathcal{T}}_2 (\bm w , \bm r, \alpha) &= \frac{g^2}{2\pi^2} \sqrt{\alpha(1-\alpha)} \left[ 
\left( - \frac{m}{w_T^2} K_0 (m r_T) \right) \bm w \right],
\end{align}
\end{subequations}
Note that the unpolarized and transversely polarized wave functions are real, while the longitudinally-polarized wave function is pure imaginary, such that
\begin{align} \label{e:WFcc}
  \left[ \tilde{\Psi}_i (\bm w, \bm r , \alpha) \right]^\dagger \equiv 
  \left[ \mathds 1 \right] \mathcal{U}_i (\bm w, \bm r , \alpha) -
  \left[ \tau_3 \right] \mathcal{L}_i  (\bm w, \bm r , \alpha) +
  \left[ \bm{\tau} \right] \times \bm{\mathcal{T}}_i   (\bm w, \bm r , \alpha) .
\end{align}

\begin{figure}
\includegraphics[width=\textwidth]{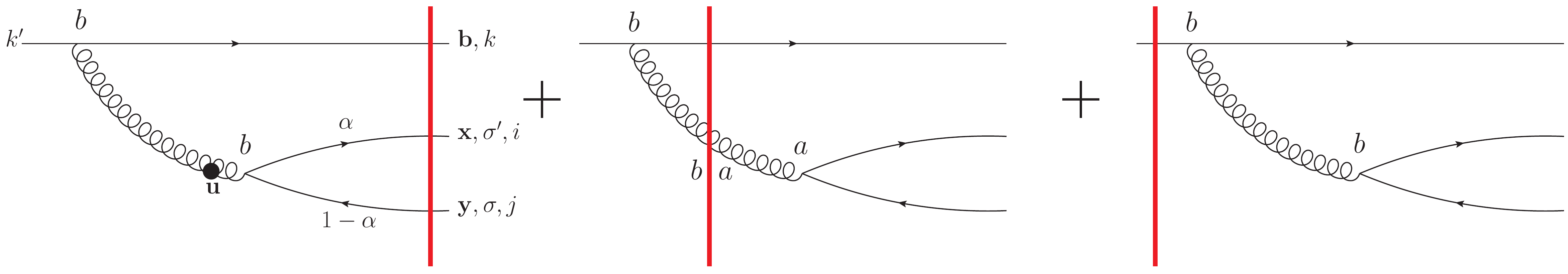}
\caption{ \label{f:buildblock}
The building block $(\mathcal{A}^{q \bar q})_{(i j) \, (k k') \, (\sigma \sigma')}$ of the scattering amplitude: a sum over time orderings for a single $q \bar q$ emission.
} 
\end{figure}

The wave function for each time ordering is subsequently dressed by the Wilson line color rotations of the partons which scatter in the field of the target nucleus at $t_{LF} = x^+ = 0$, as shown in Fig.~\ref{f:buildblock}.  The Wilson line $V_{\bm x}$ in the fundamental representation is defined as
\begin{align}
V_{\bm x} \equiv \mathcal{P} \exp\left[ i g \int dx^+ A^- (x^+, 0^-, \bm{x}) \right] ,
\end{align}
and the adjoint Wilson line $U_{\bm x}$ is defined analogously.  In the case of the three time orderings shown in Fig.~\ref{f:buildblock}, there are two color matrices built from these Wilson lines: one for the color rotation of the valence quark (with indices $k k'$) and one for the color rotation of the produced $q \bar q$ pair (with indices $i j$).  For the time-ordering $1$ in which the pair is radiated before the interaction, the wave function $\tilde{\Psi}_1$ is dressed by the color rotations $(V_{\bm b} t^b)_{k k'} \, (V_{\bm x} t^b V_{\bm y}^\dagger)_{i j}$.  For the time ordering $2$ in which the gluon is radiated before the interaction but pair produces after, the wave function $\tilde{\Psi}_2$ is dressed by the color rotations $(V_{\bm b} t^b)_{k k'} \, (U_{\bm u})^{a b} (t^a)_{i j}$.  But for the time ordering $3$ in which the valence quark scatters first and then radiates the $q \bar q$ pair after the interaction, the situation is slightly different.  The wave function $\tilde{\Psi}_3 = - \tilde{\Psi}_1 - \tilde{\Psi}_2$ is not independent of the others, and the color rotation of the valence quark is different from the other two time orderings: $(t^b V_{\bm b})_{k k'}$ vs. $(V_{\bm b} t^b)_{k k'}$.  This different color matrix for time ordering $3$ can be brought into the same form as the ones for time orderings $1$ and $2$ by using the identity $(U_{\bm b})^{a b} t^b = V_{\bm b}^\dagger t^a V_{\bm b}$ to write
%
%
%
%
\begin{align}
(t^b V_{\bm b})_{k k'} (t^b)_{i j} =  
(V_{\bm b} V_{\bm b}^\dagger t^b V_{\bm b})_{k k'} (t^b)_{i j} = 
(V_{\bm b} t^b)_{k k'} \: (U_{\bm b})^{a b}  (t^a)_{i j} ,
\end{align}
where in the last step we relabeled the dummy indices $a \leftrightarrow b$.  This trick, which is visualized in the transition of Fig.~1(d) $\rightarrow$ Fig.~3 of \cite{Mueller:2001uk}, brings all three time orderings into the same form, allowing us to write the total amplitude for a valence quark to radiate a soft $q \bar q$ pair as
\begin{align} \label{e:buildblock}
  (\tilde{\mathcal{A}}_{N A})_{(i j) \, (k k') \, (\sigma \sigma')} (\bm x , \bm y , \bm b , \bm u, \alpha)  &\equiv 
  (V_{\bm b} t^b)_{k k'} \left[ \left[ W_1^b (\bm x , \bm y , \bm b) \right]_{i j} 
  \left[ \tilde{\Psi}_1 (\bm u - \bm b , \bm x - \bm y , \alpha) \right]_{\sigma' \, , \, -\sigma} \right.
  \notag \\ & \hspace{1.75cm} +
  \left. \left[ W_2^b (\bm u , \bm b) \right]_{i j} 
  \left[ \tilde{\Psi}_2 (\bm u - \bm b , \bm x - \bm y , \alpha) \right]_{\sigma' \, , \, -\sigma} \right] ,
\end{align}
where time ordering $3$ has been absorbed into time orderings $1$ and $2$ by shifting the $q \bar q$ color rotations:
\begin{subequations} \label{e:Smatrix1}
\begin{align} 
  W_1^b (\bm x, \bm y, \bm b) &\equiv V_{\bm x} t^b V_{\bm y}^\dagger - (U_{\bm b})^{ab} t^a =
  V_{\bm x} t^b V_{\bm y}^\dagger - V_{\bm b} t^b V_{\bm b}^\dagger \\
  W_2^b (\bm u, \bm b) &\equiv (U_{\bm u})^{ab} t^a - (U_{\bm b})^{a b} t^a =
  V_{\bm u} t^b V_{\bm u}^\dagger - V_{\bm b} t^b V_{\bm b}^\dagger .
\end{align}
\end{subequations}
The fundamental building block \eqref{e:buildblock} is an operator matrix in the spins and colors of the various partons, and by squaring and tracing out these quantum numbers in the appropriate ways, we can form all contributions to the single- and double-pair production cross-sections.

%
\subsection{Single Pair Production}
%

As a first application of the building block \eqref{e:buildblock}, let us compute the single-pair production cross-section \eqref{e:CSsingle} contributing to the $q \bar q$ correlation function \eqref{e:qaqcorr1}.   Squaring the building block \eqref{e:buildblock} and averaging over the color and spin of the valence quark, we obtain 
\begin{align} \label{e:asq}
\left\langle | \tilde{\mathcal{A}}_{N A} |^2 \right\rangle &= \frac{1}{2 N_c} \sum_{i, j = 1}^2
\tr_\tau [\tilde{\Psi}_i \tilde{\Psi}_j^\dagger] \: \left\langle
\tr_C [ V_{\bm b_1} t^b t^c V_{\bm b_1}^\dagger ] \:
\tr_C [ W_i^b W_j^{c \, \dagger} ] \right\rangle
\notag \\ &=
\frac{1}{2 N_c} \sum_{i, j = 1}^2 
\Big( \mathcal{U}_i \mathcal{U}_j - \mathcal{L}_i \mathcal{L}_j + 
\bm{\mathcal T}_i \cdot \bm{\mathcal T}_j \Big)
(\bm{B}_1 - \bm{B}_2 , \bm{u} - \bm{b}_1 , \alpha)
\notag \\ & \hspace{1cm} \times
\left\langle \tr_C [ W_i^b W_j^{b \, \dagger} ] \right\rangle
(\bm{B}_1 , \bm{B}_2 , \bm{u} , \bm{b}_1 , \alpha) ,
\end{align}
where $\tr_C$ denote a trace over colors in the fundamental representation and $\tr_\tau$ a trace over spins and the remaining angle brackets denote an average over color configurations of the target
nucleus.\footnote{We note that the averaging denoted by angle brackets in \eqref{e:asq} is different from the event averaging denoted by $\langle \cdots \rangle_{ev}$ in, for example, \eqref{e:prob3}.  The full event averaging there denotes an averaging over configurations of both the projectile and the target, while the angle brackets over the Wilson lines in \eqref{e:asq} denote an averaging over the target only.  The averaging over configurations of the projectile has been handled explicitly in the cross-sections \eqref{e:CSsingle} and \eqref{e:CSdouble} through an integration over the density profile and the use of light-front wave functions.}
Note that, because we have expressed the spin structure of the wave functions in terms of Pauli matrices, the spin trace is trivial (using $\tr[\tau_\alpha \tau_\beta] = 2 \delta_{\alpha \beta}$) and can be straightforwardly generalized to more complex structures.  The nontrivial aspect of the calculation is the color algebra
\begin{align}
\Omega_{i j} &\equiv \tr_C [ W_i^b W_j^{b \, \dagger} ]
\end{align}
for each of the time orderings $i, j \in \{ 1, 2 \}$.  Inserting the Wilson line interactions \eqref{e:Smatrix1}, we obtain
\begin{subequations} \label{e:singleops}
\begin{align}
\Omega_{1 1} &= 2 N_c C_F - \half N_c^2 \left\langle \Dtwo{B_1}{b_1} \Dtwo{b_1}{B_2} \right\rangle 
- \half N_c^2 \left\langle \Dtwo{B_2}{b_1} \Dtwo{b_1}{B_1} \right\rangle 
\notag \\ & \hspace{1cm} + 
\half \left\langle \Dtwo{B_1}{B_2} \right\rangle + 
\half \left\langle \Dtwo{B_2}{B_1} \right\rangle
\\ \notag \\
\Omega_{1 2} &= \half N_c^2 + \half N_c^2 \left\langle \Dtwo{B_1}{u} \Dtwo{u}{B_2} \right\rangle 
- \half N_c^2 \left\langle \Dtwo{B_1}{b_1} \Dtwo{b_1}{B_2} \right\rangle 
\notag \\ & \hspace{1cm} -
\half N_c^2 \left\langle \Dtwo{u}{b_1} \Dtwo{b_1}{u} \right\rangle 
\\ \notag \\
\Omega_{2 1} &= \half N_c^2 + \half N_c^2 \left\langle \Dtwo {B_2}{u} \Dtwo {u}{B_1} \right\rangle - \half N_c^2 \left\langle \Dtwo {B_2}{b_1} \Dtwo {b_1}{B_1} \right\rangle 
\notag \\ & \hspace{1cm} -
\half N_c^2 \left\langle \Dtwo{u}{b_1} \Dtwo{b_1}{u} \right\rangle 
\\ \notag \\
\Omega_{2 2} &= N_c^2 - N_c^2 \left\langle \Dtwo{u}{b_1} \Dtwo{b_1}{u} \right\rangle ,
\end{align}
\end{subequations}
where $C_F = \frac{N_c^2 - 1}{2 N_c}$ is the quadratic Casimir factor and we recall that $\bm{u} \equiv \alpha \bm{B}_1 + (1-\alpha) \bm{B}_2$ with $\alpha \equiv \frac{k_1^+}{k_1^+ + k_2^+}$.  Here we have introduced the operator $\hat{D}_2$ for the quark dipole scattering amplitude,
\begin{align}
\Dtwo{x}{y} \equiv \frac{1}{N_c} \tr_C \left[ V_{\bm x} V_{\bm y}^\dagger \right].
\end{align}
As seen in \eqref{e:singleops}, the average of $\hat{D}_2$ and of products of $\hat{D}_2$ determine the interactions' contribution to the $q \bar q$ correlation function.  At higher orders, more and more complex Wilson line traces will become relevant, including the quadrupole $\hat{D}_4$, sextupole $\hat{D}_6$, and octupole $\hat{D}_8$ operators:
\begin{subequations}
\begin{align}
\Dfour{x}{y}{z}{w} &\equiv \frac{1}{N_c} \tr_C \left[ V_{\bm x} V_{\bm y}^\dagger \: 
V_{\bm z} V_{\bm w}^\dagger \right]
\\ 
\Dsix{x}{y}{z}{w}{u}{v} &\equiv \frac{1}{N_c} \tr_C \left[ V_{\bm x} V_{\bm y}^\dagger \: 
V_{\bm z} V_{\bm w}^\dagger \: V_{\bm u} V_{\bm v}^\dagger \right]
\\ 
\Deight{x}{y}{z}{w}{u}{v}{r}{s} &\equiv \frac{1}{N_c} \tr_C \left[ V_{\bm x} V_{\bm y}^\dagger \: 
V_{\bm z} V_{\bm w}^\dagger \: V_{\bm u} V_{\bm v}^\dagger \: V_{\bm r} V_{\bm s}^\dagger \right] .
\end{align}
\end{subequations}
Inserting these operators into the cross-section \eqref{e:CSsingle} gives the lowest-order expression
\begin{align} \label{e:singlecorr}
k_1^+ k_2^+ & \frac{d\sigma^{q \bar q}}{d^2 B_1 dk_1^+ \, d^2 B_2 dk_2^+} = 
\frac{a}{(4\pi)^2} \int d^2 b_1 
\notag \\ & 
\times \Bigg\{ \Big( \mathcal{U}_1 \mathcal{U}_1 - \mathcal{L}_1 \mathcal{L}_1 + 
\bm{\mathcal T}_1 \cdot \bm{\mathcal T}_1 \Big) (\bm{B}_1 - \bm{B}_2 , \bm{u} - \bm{b}_1 , \alpha)
\notag \\ & \hspace{1cm}
\times \Bigg[ C_F - \frac{N_c}{4} \left\langle \Dtwo{B_1}{b_1} \Dtwo{b_1}{B_2} \right\rangle 
- \frac{N_c}{4} \left\langle \Dtwo{B_2}{b_1} \Dtwo{b_1}{B_1} \right\rangle 
\notag \\ & \hspace{2cm}
+ \frac{1}{4N_c} \left\langle \Dtwo{B_1}{B_2} \right\rangle + 
\frac{1}{4N_c} \left\langle \Dtwo{B_2}{B_1} \right\rangle \Bigg]
\notag \\ & 
+ \Big( \mathcal{U}_1 \mathcal{U}_2 - \mathcal{L}_1 \mathcal{L}_2 + 
\bm{\mathcal T}_1 \cdot \bm{\mathcal T}_2 \Big) (\bm{B}_1 - \bm{B}_2 , \bm{u} - \bm{b}_1 , \alpha)
\notag \\ & \hspace{1cm}
\times \Bigg[ \frac{N_c}{2} - \frac{N_c}{2} \left\langle \Dtwo{u}{b_1} \Dtwo{b_1}{u} \right\rangle
+ \frac{N_c}{4} \left\langle \Dtwo{B_1}{u} \Dtwo{u}{B_2} \right\rangle 
\notag \\ & \hspace{2cm}
+ \frac{N_c}{4} \left\langle \Dtwo {B_2}{u} \Dtwo {u}{B_1} \right\rangle 
- \frac{N_c}{4} \left\langle \Dtwo{B_1}{b_1} \Dtwo{b_1}{B_2} \right\rangle 
\notag \\ &  \hspace{2cm}
- \frac{N_c}{4} \left\langle \Dtwo {B_2}{b_1} \Dtwo {b_1}{B_1} \right\rangle \Bigg]
\notag \\ &  
+ \Big( \mathcal{U}_2 \mathcal{U}_2 - \mathcal{L}_2 \mathcal{L}_2 + 
\bm{\mathcal T}_2 \cdot \bm{\mathcal T}_2 \Big) (\bm{B}_1 - \bm{B}_2 , \bm{u} - \bm{b}_1 , \alpha)
\notag \\ & \hspace{1cm}
\Bigg[ \frac{N_c}{2} - \frac{N_c}{2} \left\langle \Dtwo{u}{b_1} \Dtwo{b_1}{u} \right\rangle \Bigg] 
\Bigg\} .
\end{align}
Note that the product of the wave functions \eqref{e:WFlabels} is manifestly real and that, since $\Dtwo{x}{y}^* = \Dtwo{y}{x}$, the interactions come in pairs which are also manifestly real.  This expression is simply the coordinate-space analog of the momentum-space $q \bar q$ cross-section calculated previously \cite{Kovchegov:2006qn,Gelis:2004jp,Blaizot:2004wv,Fujii:2006ab,Gelis:2003vh,Levin:1991ry}.  By integrating out either the quark or the antiquark from \eqref{e:singlecorr} we can obtain the uncorrelated background \eqref{e:uncorr} which needs to be subtracted to obtain the correlation function \eqref{e:qaqcorr1}.  

Finally, the correlation function is normalized by the inelastic cross-section, which we take to be single-inclusive gluon production at lowest order.  The standard textbook result \cite{Kovchegov:2012mbw} is
\begin{align}
k^+ \frac{d\sigma}{d^2 k dk^+} &= \frac{a}{(2\pi)^2} \left(\frac{\alpha_s N_c}{2 \pi^2}\right) 
\int d^2 x_2 \, d^2 x_2^\prime \, d^2 x_1  \, e^{- i \bm{k} \cdot \bm{x}_{2 2'}} \, 
\frac{\bm{x}_{21} \cdot \bm{x}_{2' 1}}{x_{21}^2 \, x_{2' 1}^2} 
\notag \\ & \times
\left\langle 1 + \left| \Dtwo{x_2}{x_2^\prime} \right|^2 - \left| \Dtwo{x_1}{x_2^\prime} \right|^2 - 
\left| \Dtwo{x_2}{x_1} \right|^2 \right\rangle ,
\end{align}
with $\bm{x}_{ij} \equiv \bm{x}_i - \bm{x}_j$, and the corresponding total cross-section is
\begin{align} \label{e:siginel}
\sigma_{inel} &= \frac{\alpha_s N_c}{\pi^2} \, (a \, \Delta Y) \, \int \frac{ d^2 x_1 \, d^2 x_2}{(x_{21})_T^2}
\left\langle 1 - \left| \Dtwo{x_1}{x_2} \right|^2 \right\rangle ,
\end{align}
with $\Delta Y \equiv \int\limits_{p_A^+}^{p_a^+} \frac{dk^+}{k^+} = \ln\frac{p_a^+}{p_A^+} = \ln\frac{s}{M_A^2}$ the total rapidity interval of the collision.  Note that the scale to which the center-of-mass energy squared $s$ is compared in the rapidity is not uniquely fixed; with eikonal accuracy one could equally well define $\Delta Y = \ln\frac{s}{M_a M_A}$ instead.

Once an explicit model is chosen for the target averaging, one can calculate the correlations explicitly, using the same model for both the $q \bar q$ production and for $\sigma_{inel}$.  One simple and widely-used model which can be employed is the McLerran-Venugopalan (MV) model~\cite{McLerran:1993ni,McLerran:1993ka,McLerran:1994vd}.  The average for the dipole amplitude in the MV model is well known, but the average of products of dipole amplitudes is more esoteric.  Still, the MV model evaluation of the  double-dipole is known in the literature, with the explicit form \cite{Dominguez:2008aa} 
\begin{subequations} \label{e:MV}
\begin{align} \label{e:MV1}
\left\langle \Dtwo{B_1}{B_2} \right\rangle &= e^{-\frac{1}{4} |\bm{B}_1 - \bm{B}_2|_T^2 Q_s^2}
\\\notag \\
\left\langle \Dtwo{x}{y} \Dtwo{u}{v} \right\rangle &= e^{-\frac{1}{4} |\bm{x} - \bm{y}|_T^2 Q_s^2} \:
e^{-\frac{1}{4} |\bm{u} - \bm{v}|_T^2 Q_s^2} \: e^{-\frac{1}{4} \frac{N_c}{2 C_F} (\bm{x} - \bm{u}) \cdot (\bm{y} - \bm{v}) Q_s^2}
\notag \\ & \hspace{-3cm} \times
\Bigg[
\Bigg(\frac{(\bm{x}-\bm{u}) \cdot (\bm{y} - \bm{v}) + (\tfrac{2 C_F}{Q_s^2}) \mu^2 \sqrt{\Delta}}
{2 \: (\tfrac{2 C_F}{Q_s^2}) \mu^2 \sqrt{\Delta}} - \frac{1}{N_c^2} \frac{(\bm{x}-\bm{y}) \cdot (\bm{u} - \bm{v})}
{(\tfrac{2 C_F}{Q_s^2}) \mu^2 \sqrt{\Delta}} \Bigg) \: e^{+\tfrac{1}{4} \tfrac{N_c}{2 C_F} (\tfrac{2 C_F}{Q_s^2}) \mu^2 \sqrt{\Delta} Q_s^2} 
\notag \\ & \hspace{-3cm} -
\Bigg(\frac{(\bm{x}-\bm{u}) \cdot (\bm{y} - \bm{v}) - (\tfrac{2 C_F}{Q_s^2}) \mu^2 \sqrt{\Delta}}
{2 \: (\tfrac{2 C_F}{Q_s^2}) \mu^2 \sqrt{\Delta}} - \frac{1}{N_c^2} \frac{(\bm{x}-\bm{y}) \cdot (\bm{u} - \bm{v})}
{(\tfrac{2 C_F}{Q_s^2}) \mu^2 \sqrt{\Delta}} \Bigg) \: e^{-\tfrac{1}{4} \tfrac{N_c}{2 C_F} (\tfrac{2 C_F}{Q_s^2}) \mu^2 \sqrt{\Delta} Q_s^2} \Bigg]
\\\notag \\
\left(\frac{2 C_F}{Q_s^2}\right) \, \mu^2 \sqrt{\Delta} &= \sqrt{
\left[ (\bm{x} - \bm{u}) \cdot (\bm{y} - \bm{v}) \right]^2 + \tfrac{4}{N_c^2} 
\left[ (\bm{x} - \bm{y}) \cdot (\bm{u} - \bm{v}) \right] 
\left[ (\bm{x} - \bm{v}) \cdot (\bm{u} - \bm{y}) \right] }
\end{align}
\end{subequations}
These expressions are valid at finite $N_c$ (that is, without taking the large-$N_c$ limit).  In general, these expressions may contain an additional logarithm $\ln\frac{1}{r_T \Lambda}$ which recovers the recover the perturbative power-law tail at asymptotically large transverse momentum (and hence asymptotically short distances); that logarithm has been neglected in these expressions from \cite{Dominguez:2008aa}.  This is also known as the Gaussian approximation or the Golec-Biernat--Wusthoff (GBW) model~\cite{GolecBiernat:1998js}.

%
\subsection{Double Pair Production: Wave Functions}
%

%
\begin{figure}
\includegraphics[width=0.7\textwidth]{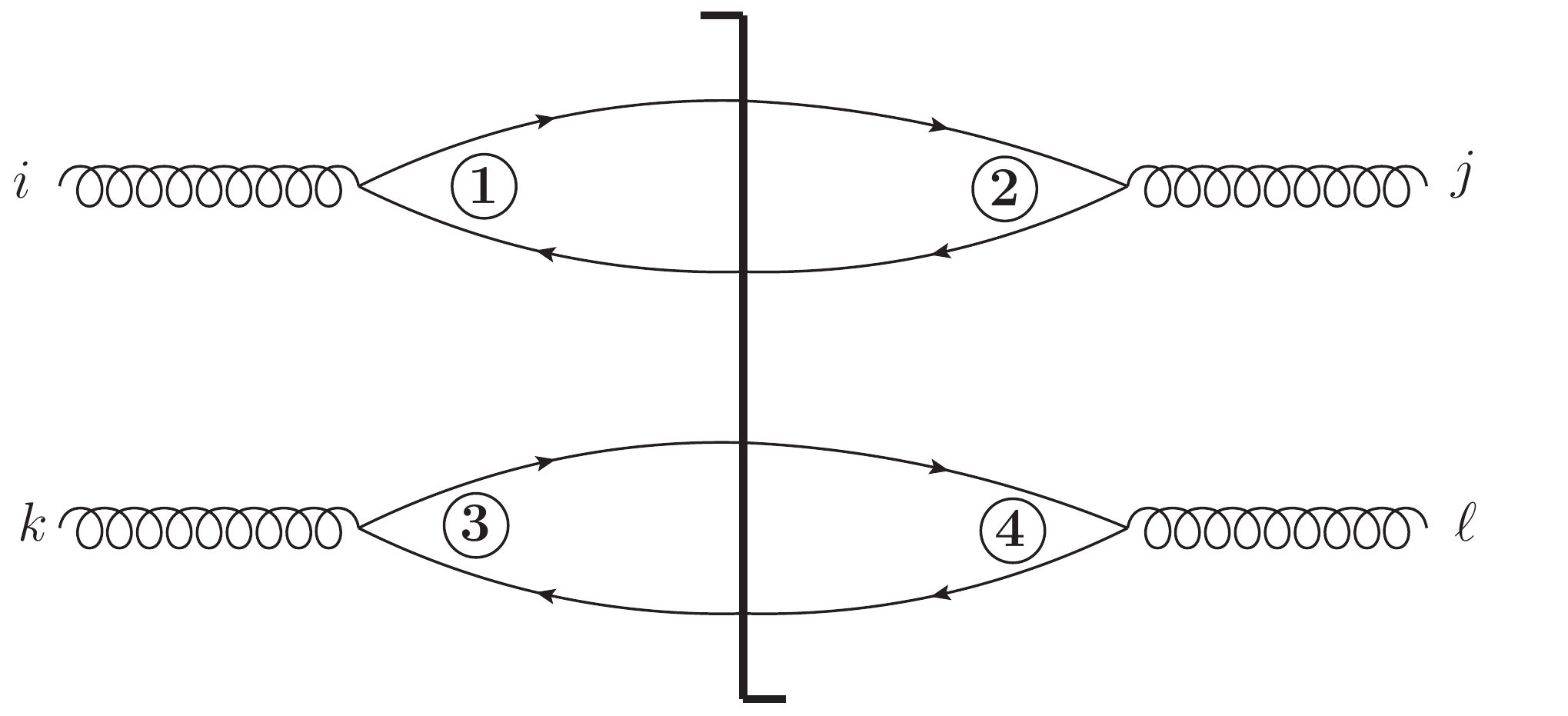}
\caption{ \label{f:Sausage_WF}
Wave functions $\Psi_{i j k \ell}^{(pairs)}$ for topologies without fermion entanglement, which form two independent fermion loops.  The source gluons are kept free to be contracted in all possible ways, and the free indices $i, j, k, \ell \in \{1, 2\}$ denote possible time orderings for each pair emission.
} 
\end{figure}
%

The calculation of the double-pair production cross-section \eqref{e:CSdouble} is enormously more complex than the calculation of the single-pair production cross-section \eqref{e:CSsingle}, for many reasons.  One is that, when two $q \bar q$ pairs have been produced, the pairs can become entangled in highly nontrivial ways: between the amplitude and complex-conjugate amplitude, the pairs may swap ownership of the quark (``quark entanglement''), the antiquark (``antiquark entanglement''), or neither (``no fermion entanglement'').  This flow of the $q \bar q$ fermion loop determines the structure of the spin trace $\tr_\tau$ over the wave functions, with each topology leading to a different wave-function structure.  

The simplest case for the wave functions in double-pair production is the case of no fermion entanglement, as shown in Fig.~\ref{f:Sausage_WF}.  Each pair is radiated from a valence quark in the light nucleus with a particular time ordering: $i, k \in \{1,2\}$ in the amplitude and $j, \ell \in \{1,2\}$ in the complex-conjugate amplitude.  This reflects another source of the greatly increased complexity for the double-pair case: now there are $2^4 = 16$ different combinations to consider (for each channel) due to the various time orderings, compared to $2^2 = 4$ combinations in the single-pair case.

Each pair is also produced with a set of kinematic arguments: the positions of the quark and antiquark, the center-of-momentum position of the pair corresponding to the position of the gluon, the position of the valence quark which radiated the pair, and the fraction of the pair momentum carried by the quark.  The details of these kinematics for the various pairs will depend on the precise diagram, but for our purposes it is convenient to denote the set of these kinematic arguments collectively by $\oone$, $\otwo$, $\othree$, and $\ofour$.  

With this compact notation shown in Fig.~\ref{f:Sausage_WF}, we can perform the spin trace $\tr_\tau$ on the pair wave functions to determine the spatial structure associated with this channel:
\begin{align} \label{e:WFpairs}
  \Psi_{i j k \ell}^{( pairs )} (\oone , \otwo, \othree, \ofour) &\equiv
  \tr_\tau \Big[ \tilde{\Psi}_{i \oone} \tilde{\Psi}^\dagger_{j \otwo} \Big]
  \tr_\tau \Big[ \tilde{\Psi}_{k \othree} \tilde{\Psi}^\dagger_{\ell \ofour} \Big]
  \notag \\ &=
  4 \Big( \mathcal{U}_{i \oone} \mathcal{U}_{j \otwo} - \mathcal{L}_{i \oone} \mathcal{L}_{j \otwo}
  + \bm{\mathcal{T}}_{i \oone} \cdot \bm{\mathcal{T}}_{j \otwo} \Big)
  \notag \\ &\hspace{1cm} \times
  \Big( \mathcal{U}_{k \othree} \mathcal{U}_{\ell \ofour} - \mathcal{L}_{k \othree} \mathcal{L}_{\ell \ofour}
  + \bm{\mathcal{T}}_{k \othree} \cdot \bm{\mathcal{T}}_{\ell \ofour} \Big) ,
\end{align}
with $i, j, k, \ell \in \{1, 2\}$.  In this simplest case, the fermion flow simply factorizes into two bubbles, giving the square of the single-pair production case.  Likewise, the color traces over the Wilson lines will also factorize into a product of two traces, but the precise color flow will depend on which valence quark sources each pair.

%
\begin{figure}
\includegraphics[width=0.85\textwidth]{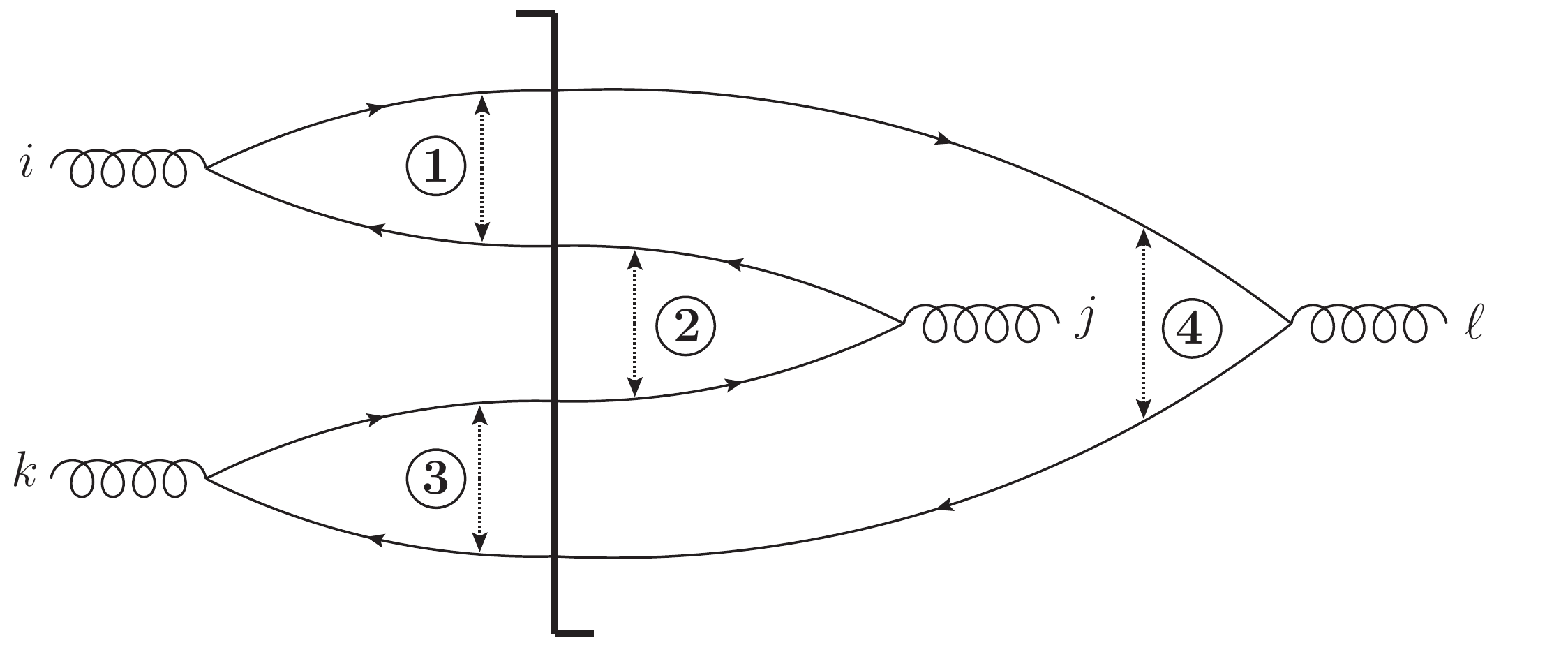}
\caption{ \label{f:WFPacMan}
Wave functions $\Psi_{i j k \ell}^{(loop)}$ for fermion entanglement topologies which form a single fermion loop.  The source gluons are kept free to be contracted in all possible ways, and the free indices $i, j, k, \ell \in \{1, 2\}$ denote possible time orderings for each pair emission.
} 
\end{figure}
%

A significantly more complicated wave function structure arises when the fermion flow is entangled between the pairs, as illustrated in Fig.~\ref{f:WFPacMan}.  Here, instead of forming two independent loops, the fermion flow is connected into one larger loop.  Depending on which valence quark radiated which pair, this may correspond to the entanglement of either the quark or the antiquark.  Using the same compact notation, we can write the wave function structure for this topology as
\newpage
\begin{align}
  \Psi_{i, j, k, \ell}^{(loop)} (\oone, \otwo, \othree, \ofour) &=
  \tr_\tau \left[ \tilde{\Psi}_{i \oone} \tilde{\Psi}^\dagger_{j \otwo} 
  \tilde{\Psi}_{k \othree} \tilde{\Psi}^\dagger_{\ell \ofour} \right] .
  \notag \\ &= 
  \tr \Big[ \left( \left[ \mathds 1 \right] \mathcal{U}_{i \oone} + \left[ \tau_3 \right] \mathcal{L}_{i \oone} +
  \left[ \bm{\tau} \right] \times \bm{\mathcal{T}}_{i \oone}  \right) 
  \notag \\ & \hspace{0.57cm} \times
  \left( \left[ \mathds 1 \right] \mathcal{U}_{j \otwo} - \left[ \tau_3 \right] \mathcal{L}_{j \otwo} +
  \left[ \bm{\tau} \right] \times \bm{\mathcal{T}}_{j \otwo}  \right)
  \notag \\ & \hspace{0.57cm} \times
  \left( \left[ \mathds 1 \right] \mathcal{U}_{k \othree} + \left[ \tau_3 \right] \mathcal{L}_{k \othree} +
  \left[ \bm{\tau} \right] \times \bm{\mathcal{T}}_{k \othree}  \right)
  \notag \\ & \hspace{0.57cm} \times
  \left( \left[ \mathds 1 \right] \mathcal{U}_{\ell \ofour} - \left[ \tau_3 \right] \mathcal{L}_{\ell \ofour} +
  \left[ \bm{\tau} \right] \times \bm{\mathcal{T}}_{\ell \ofour}  \right) \Big] .
\end{align}
This time, the Pauli trace contains four matrices, so several new structures are generated according to the algebraic properties given in \eqref{e:Pauli} of Appendix \ref{app:Pauli}.  Nontrivial results occur for traces of two, three, and four Pauli matrices, but although this result is significantly more complex, its calculation in the Pauli matrix notation remains straightforward:
\newpage
\begin{align} \label{e:WFloop}
  \Psi_{i, j, k, \ell}^{(loop)} (\oone , \otwo, \othree, \ofour) &= 
  2 \mathcal{U}_{i \oone} \mathcal{U}_{j \otwo} \mathcal{U}_{k \othree} \mathcal{U}_{\ell \ofour}  - 
  2 \mathcal{U}_{i \oone} \mathcal{U}_{j \otwo} \mathcal{L}_{k \othree} \mathcal{L}_{\ell \ofour} +
  2 \mathcal{U}_{i \oone} \mathcal{U}_{j \otwo} \bm{\mathcal{T}}_{k \othree} \cdot \bm{\mathcal{T}}_{\ell \ofour}
  \notag \\ & \hspace{-2.5cm} +
  2 \mathcal{U}_{i \oone} \mathcal{L}_{j \otwo} \mathcal{U}_{k \othree} \mathcal{L}_{\ell \ofour} -
  2 \mathcal{U}_{i \oone} \mathcal{L}_{j \otwo} \mathcal{L}_{k \othree} \mathcal{U}_{\ell \ofour} -
  2 i \mathcal{U}_{i \oone} \mathcal{L}_{j \otwo} \bm{\mathcal{T}}_{k \othree} \times \bm{\mathcal{T}}_{\ell \ofour}
  \notag \\ & \hspace{-2.5cm} +
  2 \mathcal{U}_{i \oone} \mathcal{U}_{k \othree} \bm{\mathcal{T}}_{j \otwo} \cdot \bm{\mathcal{T}}_{\ell \ofour} -
  2 i \mathcal{U}_{i \oone} \mathcal{L}_{k \othree} \bm{\mathcal{T}}_{j \otwo} \times \bm{\mathcal{T}}_{\ell \ofour} +
  2 \mathcal{U}_{i \oone} \mathcal{U}_{\ell \ofour} \bm{\mathcal{T}}_{j \otwo} \cdot \bm{\mathcal{T}}_{k \othree} 
  \notag \\ & \hspace{-2.5cm} -
  2 i \mathcal{U}_{i \oone} \mathcal{L}_{\ell \ofour} \bm{\mathcal{T}}_{j \otwo} \times \bm{\mathcal{T}}_{k \othree} -
  2 \mathcal{L}_{i \oone} \mathcal{U}_{j \otwo} \mathcal{U}_{k \othree} \mathcal{L}_{\ell \ofour} +
  2 \mathcal{L}_{i \oone} \mathcal{U}_{j \otwo} \mathcal{L}_{k \othree} \mathcal{U}_{\ell \ofour}
  \notag \\ & \hspace{-2.5cm} +
  2 i \mathcal{L}_{i \oone} \mathcal{U}_{j \otwo} \bm{\mathcal{T}}_{k \othree} \times \bm{\mathcal{T}}_{\ell \ofour} -
  2 \mathcal{L}_{i \oone} \mathcal{L}_{j \otwo} \mathcal{U}_{k \othree} \mathcal{U}_{\ell \ofour} +
  2 \mathcal{L}_{i \oone} \mathcal{L}_{j \otwo} \mathcal{L}_{k \othree} \mathcal{L}_{\ell \ofour}
  \notag \\ & \hspace{-2.5cm} -
  2 \mathcal{L}_{i \oone} \mathcal{L}_{j \otwo} \bm{\mathcal{T}}_{k \othree} \cdot \bm{\mathcal{T}}_{\ell \ofour} +
  2 i \mathcal{L}_{i \oone} \mathcal{U}_{k \othree} \bm{\mathcal{T}}_{j \otwo} \times \bm{\mathcal{T}}_{\ell \ofour} -
  2 \mathcal{L}_{i \oone} \mathcal{L}_{k \othree} \bm{\mathcal{T}}_{j \otwo} \cdot \bm{\mathcal{T}}_{\ell \ofour}
  \notag \\ & \hspace{-2.5cm} +
  2 i \mathcal{L}_{i \oone} \mathcal{U}_{\ell \ofour} \bm{\mathcal{T}}_{j \otwo} \times \bm{\mathcal{T}}_{k \othree} -
  2 \mathcal{L}_{i \oone} \mathcal{L}_{\ell \ofour} \bm{\mathcal{T}}_{j \otwo} \cdot \bm{\mathcal{T}}_{k \othree} +
  2 \mathcal{U}_{j \otwo} \mathcal{U}_{k \othree} \bm{\mathcal{T}}_{i \oone} \cdot \bm{\mathcal{T}}_{\ell \ofour}
  \notag \\ & \hspace{-2.5cm} -
  2 i \mathcal{U}_{j \otwo} \mathcal{L}_{k \othree} \bm{\mathcal{T}}_{i \oone} \times \bm{\mathcal{T}}_{\ell \ofour} +
  2 \mathcal{U}_{j \otwo} \mathcal{U}_{\ell \ofour} \bm{\mathcal{T}}_{i \oone} \cdot \bm{\mathcal{T}}_{k \othree} -
  2 i \mathcal{U}_{j \otwo} \mathcal{L}_{\ell \ofour} \bm{\mathcal{T}}_{i \oone} \times \bm{\mathcal{T}}_{k \othree} 
  \notag \\ & \hspace{-2.5cm} +
  2 i \mathcal{L}_{j \otwo} \mathcal{U}_{k \othree} \bm{\mathcal{T}}_{i \oone} \times \bm{\mathcal{T}}_{\ell \ofour} -
  2 \mathcal{L}_{j \otwo} \mathcal{L}_{k \othree} \bm{\mathcal{T}}_{i \oone} \cdot \bm{\mathcal{T}}_{\ell \ofour} +
  2 i \mathcal{L}_{j \otwo} \mathcal{U}_{\ell \ofour} \bm{\mathcal{T}}_{i \oone} \times \bm{\mathcal{T}}_{k \othree} 
  \notag \\ & \hspace{-2.5cm} -
  2 \mathcal{L}_{j \otwo} \mathcal{L}_{\ell \ofour} \bm{\mathcal{T}}_{i \oone} \cdot \bm{\mathcal{T}}_{k \othree} +
  2 \mathcal{U}_{k \othree} \mathcal{U}_{\ell \ofour} \bm{\mathcal{T}}_{i \oone} \cdot \bm{\mathcal{T}}_{j \otwo} -
  2 i \mathcal{U}_{k \othree} \mathcal{L}_{\ell \ofour} \bm{\mathcal{T}}_{i \oone} \times \bm{\mathcal{T}}_{j \otwo} 
  \notag \\ & \hspace{-2.5cm} +
  2 i \mathcal{L}_{k \othree} \mathcal{U}_{\ell \ofour} \bm{\mathcal{T}}_{i \oone} \times \bm{\mathcal{T}}_{j \otwo} -
  2 \mathcal{L}_{k \othree} \mathcal{L}_{\ell \ofour} \bm{\mathcal{T}}_{i \oone} \cdot \bm{\mathcal{T}}_{j \otwo} +
  2 (\bm{\mathcal{T}}_{i \oone} \cdot \bm{\mathcal{T}}_{j \otwo}) 
              (\bm{\mathcal{T}}_{k \othree} \cdot \bm{\mathcal{T}}_{\ell \ofour})
  \notag \\ & \hspace{-2.5cm} -
  2 (\bm{\mathcal{T}}_{i \oone} \cdot \bm{\mathcal{T}}_{k \othree}) 
              (\bm{\mathcal{T}}_{j \otwo} \cdot \bm{\mathcal{T}}_{\ell \ofour}) +
  2 (\bm{\mathcal{T}}_{i \oone} \cdot \bm{\mathcal{T}}_{\ell \ofour}) 
              (\bm{\mathcal{T}}_{j \otwo} \cdot \bm{\mathcal{T}}_{k \othree}) .
\end{align}
Note that, as $\mathcal{L}_i$ is purely imaginary, the wave function trace $\Psi_{i j k \ell}^{(loop)}$ is completely real.

%
\begin{figure}
\includegraphics[width=\textwidth]{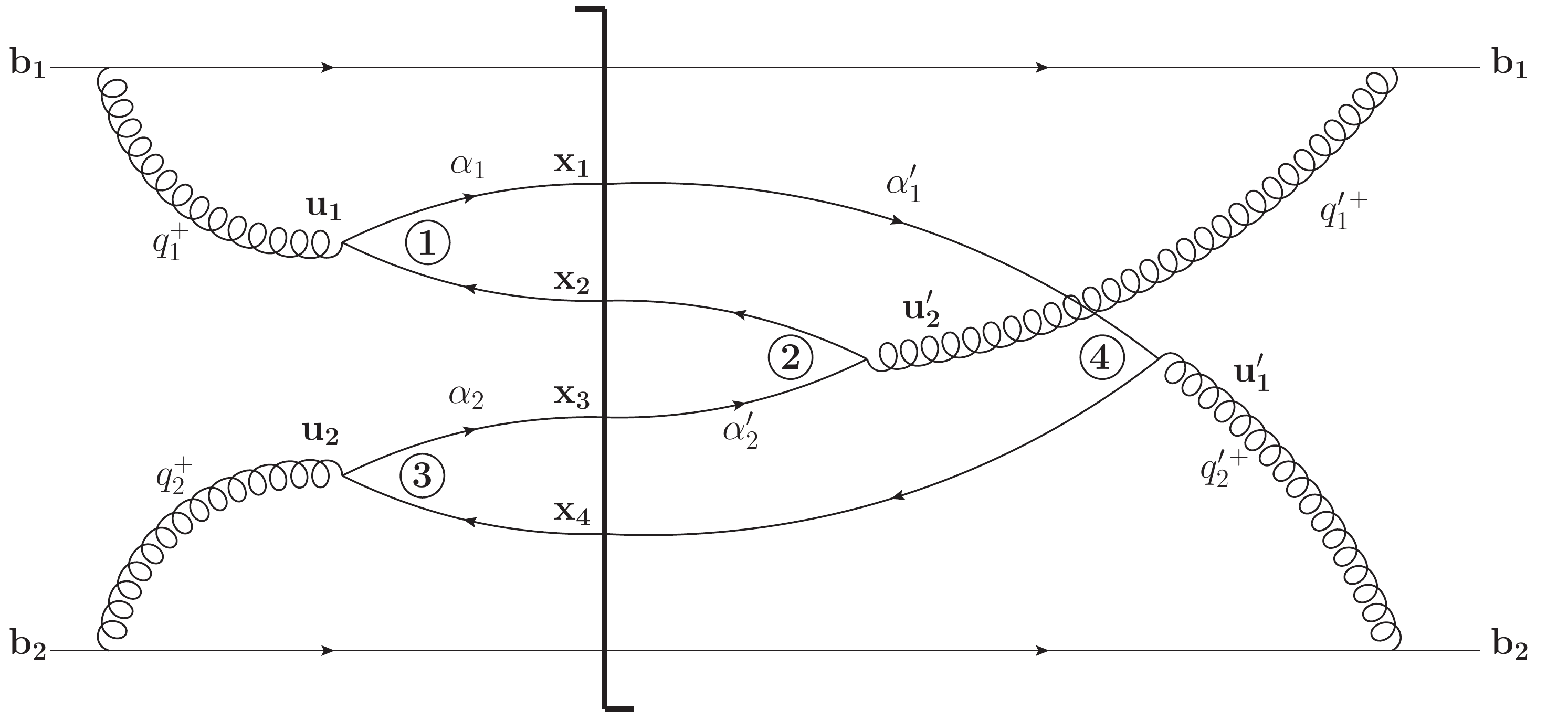}
\caption{ \label{f:qent} Quark entanglement channel of double-pair production.} 
\end{figure}
%

For these fermion entanglement topologies, the Wilson line color traces will also be arranged into a single large trace.  Depending on which valence quark radiates which pair, there are multiple color flows which can be constructed, and as a result of the highly-entangled structure, each of these color structures will generate very complex Wilson line multipoles.  To illustrate the ingredients and the complexity of such channels, let us consider the specific example of quark entanglement shown in Fig.~\ref{f:qent}.  Immediately, the entanglement of the pairs leads to an entanglement of their kinematic arguments: the momentum fractions of the pairs become scrambled, such that 
\begin{subequations}
\begin{align}
\alpha_1 = \frac{k_1^+}{k_1^+ + k_2^+} 
\hspace{1cm} &\neq \hspace{1cm}
\alpha_1^\prime = \frac{k_1^+}{k_1^+ + k_4^+} \\
\alpha_2 = \frac{k_3^+}{k_3^+ + k_4^+} 
\hspace{1cm} &\neq \hspace{1cm}
\alpha_2^\prime = \frac{k_3^+}{k_3^+ + k_2^+} ,
\end{align}
\end{subequations}
which subsequently scrambles the positions of the gluons lying at the pairs' centers of momentum:
\begin{subequations}
\begin{align}
  \bm u_1 = \alpha_1 \bm x_1 + (1-\alpha_1) \bm x_2 
  \hspace{1cm} &\neq \hspace{1cm}
  \bm u_1^\prime = \alpha_1^\prime \bm x_1 + (1-\alpha_1^\prime) \bm x_4 \\
  \bm u_2 = \alpha_2 \bm x_3 + (1-\alpha_2) \bm x_4
  \hspace{1cm} &\neq \hspace{1cm}
  \bm u_2^\prime = \alpha_2^\prime \bm x_3 + (1-\alpha_2^\prime) \bm x_2 .
\end{align}
\end{subequations}
With these definitions, the abbreviated notation for the various pairs can be spelled out explicitly as
\begin{subequations}
\begin{align}
  \oone &\equiv \{ \bm x_1 , \bm x_2 , \bm b_1 , \bm u_1 , \alpha_1 \} \\
  \otwo &\equiv \{ \bm x_3 , \bm x_2 , \bm b_1 , \bm u_2^\prime , \alpha_2^\prime \} \\
  \othree &\equiv \{ \bm x_3 , \bm x_4 , \bm b_2 , \bm u_2 , \alpha_2 \} \\
  \ofour &\equiv \{ \bm x_1 , \bm x_4 , \bm b_2 , \bm u_1^\prime , \alpha_1^\prime \} ,
\end{align}
\end{subequations}
where the arguments denote the position of the quark, antiquark, emitting valence quark, gluon, and quark momentum fraction, respectively, for each pair.

For the quark entanglement channel of double-pair production, the complete color trace is given in terms of the building block \eqref{e:Smatrix1} by
\begin{align}
  \Omega_{i j k \ell}^{(q \: ent)} (\oone, \otwo, \othree, \ofour) &\equiv 
  \tr\left[ W_{i \oone}^b W_{j \otwo}^{\dagger \, b} \, W_{k \othree}^c W_{\ell \ofour}^{\dagger \, c} \right] .
\end{align}
There are $2^4 = 16$ such terms corresponding to the different time orderings $i, j, k, \ell \in \{1, 2\}$ for this channel alone, and each of these color traces is quite long and generally involves high-order Wilson line multipoles.  To illustrate this, consider just the first contribution $i = j = k = \ell = 1$, for which the explicit interaction is given by
\begin{align} \label{e:intexample}
4\Omega_{1111}^{(q \: ent)} &= 
8 C_F^2 N_c - 2 C_F N_c^2 \Dtwo{b_1}{x_2} \Dtwo{x_1}{b_1} - 2 C_F N_c^2 \Dtwo{b_2}{x_4} \Dtwo{x_1}{b_2} 
\notag \\ &
+ 4 C_F^2 N_c \Dtwo{x_1}{x_3} - 2 C_F N_c^2 \Dtwo{b_1}{x_1} \Dtwo{x_2}{b_1} 
\notag \\ &
- 2 C_F N_c^2 \Dtwo{b_1}{x_3} \Dtwo{x_2}{b_1} + N_c^3 \Dtwo{b_1}{b_2} \Dtwo{b_2}{x_4} \Dtwo{x_2}{b_1} 
\notag \\ &
- 2 C_F N_c^2 \Dtwo{b_1}{x_2} \Dtwo{x_3}{b_1} - 2 C_F N_c^2 \Dtwo{b_2}{x_4} \Dtwo{x_3}{b_2} 
\notag \\ &
+ 4 C_F^2 N_c \Dtwo{x_3}{x_1} + N_c^3 \Dtwo{b_1}{x_2} \Dtwo{b_2}{b_1} \Dtwo{x_4}{b_2} 
\notag \\ &
- 2 C_F N_c^2 \Dtwo{b_2}{x_1} \Dtwo{x_4}{b_2} - 2 C_F N_c^2 \Dtwo{b_2}{x_3} \Dtwo{x_4}{b_2} 
\notag \\ &
+ N_c^3 \Dtwo{x_2}{b_1} \Dtwo{x_4}{b_2} \Dfour{b_1}{x_3}{b_2}{x_1} 
\notag \\ &
+ N_c^3 \Dtwo{b_1}{x_2} \Dtwo{b_2}{x_4} \Dfour{x_1}{b_1}{x_3}{b_2}
\notag \\ \notag \\ &+
2 C_F \Dtwo{x_1}{x_2} + 2 C_F \Dtwo{x_1}{x_4} - N_c \Dtwo{b_1}{x_4} \Dtwo{x_2}{b_1} 
\notag \\ &
- N_c \Dtwo{b_2}{x_4} \Dtwo{x_2}{b_2} + 2 C_F \Dtwo{x_2}{x_1} + 2 C_F \Dtwo{x_2}{x_3} + 2 C_F \Dtwo{x_3}{x_2} 
\notag \\ &
+ 2 C_F \Dtwo{x_3}{x_4} - N_c \Dtwo{b_1}{x_2} \Dtwo{x_4}{b_1} - N_c \Dtwo{b_2}{x_2} \Dtwo{x_4}{b_2} 
\notag \\ &
+ 2 C_F \Dtwo{x_4}{x_1} + 2 C_F \Dtwo{x_4}{x_3} - N_c \Dtwo{x_2}{b_1} \Dfour{b_1}{x_3}{x_4}{x_1} 
\notag \\ &
- N_c \Dtwo{b_1}{x_2} \Dfour{x_1}{b_1}{x_3}{x_4} - N_c \Dtwo{b_2}{x_4} \Dfour{x_1}{x_2}{x_3}{b_2} 
\notag \\ &
- N_c \Dtwo{x_4}{b_2} \Dfour{x_2}{x_3}{b_2}{x_1}
\notag \\ \notag \\ &+
\frac{1}{N_c} \Dtwo{x_2}{x_4} + \frac{1}{N_c} \Dtwo{x_4}{x_2} + \frac{1}{N_c} \Dfour{x_1}{x_2}{x_3}{x_4} 
\notag \\ &
+ \frac{1}{N_c} \Dfour{x_2}{x_3}{x_4}{x_1} .
\end{align}
This operator, which must yet be averaged over color configurations of the target, involves products of one, two, and three dipoles, quadrupoles, and the product of one or two dipoles with a quadrupole.  Clearly, an expression of this complexity will be difficult to compute even in a simple analytic model like the MV / GBW model, and some resort to approximate or numerical techniques will be necessary.  For this reason, we defer a detailed analysis of all the ensuing color structures to a dedicated future publication, in which we will study aspects of these full correlations analytically and numerically.  Still, there are general physical features of the correlation functions we can understand by studying the single-pair production result \eqref{e:singlecorr} and the wave functions \eqref{e:WFpairs} and \eqref{e:WFloop}, which we will pursue next.

%
\section{Correlations over Various Length Scales}
\label{sec:lengths}
%

%
\subsection{Single-Pair Production: Long-Distance Asymptotics}
%

One important quantification of the various correlation functions $\mathcal{C}_{i j}$ is their characteristic correlation length $L_{i j}$ which controls the exponential falloff of the correlation function at large distances $|\bm{B}_1 - \bm{B}_2|_T \rightarrow \infty$:
\begin{align}
 \mathcal{C}_{i j} (\bm{B}_1 , k_1^+ \, ; \, \bm{B}_2 , k_2^+) \sim e^{- \frac{| \bm{B}_1 - \bm{B}_2 |_T}{L_{i j}}} .
\end{align}
To understand the physical picture embodied by the correlation functions $\mathcal{C}_{i j}$, let us compute their long-distance asymptotics.  

Let us start by focusing on the single-pair production cross-section \eqref{e:CSsingle}.  This cross-section only generates one $q \bar q$ pair, and as such can only contribute to the quark-antiquark correlation function $\mathcal{C}_{q \bar q}$ given in \eqref{e:singlecorr}.  As seen clearly in \eqref{e:singlecorr}, there are two sources of spatial dependence: the wave functions $\mathcal{U}_i$ , $\mathcal{L}_i$, and $\bm{\mathcal{T}}_i$ which are given in \eqref{e:WFlabels} and the interactions.  The interactions can be modeled in different ways, but for definiteness let us consider the expressions in the MV / GBW model given in \eqref{e:MV}.  Many of these interactions are exponentially suppressed at large distances due to dynamical color screening effects embodied in the saturation scale $Q_s$.  The single dipole scattering amplitude, for instance, is
\begin{align}
\left\langle \Dtwo{B_1}{B_2} \right\rangle = \left\langle \Dtwo{B_2}{B_1} \right\rangle =
e^{-\frac{1}{4} |\bm{B}_1 - \bm{B}_2|_T^2 Q_s^2} ,
\end{align}
and the majority of the double-dipole amplitudes have the same long-distance asymptotics:
\begin{align}
\langle \Dtwo{B_1}{b_1} \Dtwo{b_1}{B_2} \rangle &=
\langle \Dtwo {B_2}{b_1} \Dtwo{b_1}{B_1} \rangle
\notag \\ & \hspace{-2cm}
\sim \langle \Dtwo{B_1}{u} \Dtwo{u}{B_2} \rangle 
= \langle \Dtwo {B_2}{u} \Dtwo{u}{B_1} \rangle 
\notag \\ &
\sim \left(1 - \frac{1}{N_c^2}\right) 
e^{-\frac{1}{4} \: [\alpha^2 + (1-\alpha)^2 - \frac{1}{N_c C_F} \alpha (1-\alpha) ] \: |\bm{B}_1 - \bm{B}_2|_T^2 Q_s^2} .
\end{align}
Thus we immediately see that the full structure of the single-pair $\mathcal{C}_{q \bar q}$ correlation exists on length scales $|\bm{B}_1 - \bm{B}_2|_T \leq 1 / Q_s$.  On length scales larger than $1/Q_s$, these terms die off, but the overall correlation function $\mathcal{C}_{q \bar q}$ is not similarly suppressed.  In addition to the constant terms from the interactions, which are obviously unsuppressed for $|\bm{B}_1 - \bm{B}_2|_T \gg 1 / Q_s$, there is also the nontrivial double-dipole amplitude 
\begin{align}
\left\langle \Dtwo{u}{b_1} \Dtwo{b_1}{u} \right\rangle &= 
1 - \left(\frac{2 C_F}{N_c} \right) \left[ 1 - e^{-\frac{1}{4} \frac{N_c}{C_F} |\bm{u} - \bm{b_1}|_T^2 Q_s^2} \right] .
\end{align}
Thus for $|\bm{B}_1 - \bm{B}_2|_T \gg 1 / Q_s$ in the MV / GBW model, the $q \bar q$ cross-section is given by the simplified form
\begin{align}
k_1^+ k_2^+ \frac{d\sigma^{q \bar q}}{d^2 B_1 dk_1^+ \, d^2 B_2 dk_2^+} &\sim 
\frac{2 a C_F}{(4\pi)^2} \int d^2 b_1 \Bigg\{ 
\Big[ |\mathcal{U}_1|^2 + |\mathcal{L}_1|^2 + |\mathcal{T}_1|_T^2 \Big] 
(\bm{B}_1 - \bm{B}_2 , \bm{u} - \bm{b_1} , \alpha)
\notag \\ & \hspace{-1cm} +
\Big[ \mathcal{U}_1 \mathcal{U}_2 - \mathcal{L}_1 \mathcal{L}_2 + \bm{\mathcal{T}_1} \cdot \bm{\mathcal{T}_2}
+ |\mathcal{U}_2|^2 + |\mathcal{L}_2|^2 + |\mathcal{T}_2|_T^2 \Big] 
(\bm{B}_1 - \bm{B}_2 , \bm{u} - \bm{b_1} , \alpha)
\notag \\ & \times
\left[ 1 - e^{-\frac{1}{4} \frac{N_c}{C_F} |\bm{u} - \bm{b_1}|_T^2 Q_s^2} \right] \Bigg\} .
\end{align}

This analysis shows that there are two interesting regimes of $\mathcal{C}_{q \bar q}$ in the case of single-pair production: $|\bm{B}_1 - \bm{B}_2|_T \leq 1 / Q_s$ and $|\bm{B}_1 - \bm{B}_2|_T \gg 1 / Q_s$.  The former is sensitive to very short-distance correlations driven by the interactions, while most of these effects have died away for the latter.  This also shows that $Q_s$ does not set the overall correlation length $L_{q \bar q}$, since the correlation function overall does not decay exponentially when $|\bm{B}_1 - \bm{B}_2|_T \gg 1 / Q_s$.  The correlation length is instead set by the long-distance asymptotics of the wave functions, which are insensitive to $Q_s$.  

To determine the overall correlation length $L_{q \bar q}$, then, we need to compute the long-distance asymptotics of the wave functions and the integrals $F_0 , F_1 , F_2$ given in \eqref{e:Fdefs} from which they are built.  In the regime $m r_T \gg 1$, the asymptotics of these integrals are
\begin{subequations}
\begin{align}
F_0 (w_T , r_T , \alpha) &\approx \sqrt{\frac{\pi}{2}} \frac{\sqrt{m r_T}}{\alpha (1-\alpha) r_T^2} 
e^{- m r_T \left( 1 + \frac{w_T^2}{2 \alpha (1-\alpha) r_T^2} \right)} 
\\
F_1 (w_T , r_T , \alpha) &\approx \sqrt{\frac{\pi}{2}} 
\frac{1}{\sqrt{m r_T}} \frac{1}{w_T} e^{- m r_T} 
\left[ 1 - e^{- \frac{m r_T \, w_T^2}{2 \alpha (1-\alpha) r_T^2} } \right]
\\
F_2 (w_T , r_T , \alpha) &\approx \sqrt{\frac{\pi}{2}} 
\sqrt{m r_T} \frac{1}{r_T w_T} e^{- m r_T} 
\left[ 1 - e^{- \frac{m r_T \, w_T^2}{2 \alpha (1-\alpha) r_T^2} } \right] .
\end{align}
\end{subequations}
Propagating this forward to the wave functions, we find that all of the terms $\mathcal{U}_1 , \mathcal{U}_2 , \mathcal{L}_1 , \mathcal{L}_2 , \bm{\mathcal{T}}_1 , \bm{\mathcal{T}}_2$ at long distances all have the same limiting behavior:
\begin{align}
\mathcal{U}_i \, , \, \mathcal{L}_i \, , \, \bm{\mathcal{T}}_i \propto e^{- m |\bm{B}_1 - \bm{B}_2|_T}.
\end{align}
The wave function squared and hence the single-pair contribution to the correlation function $\mathcal{C}_{q \bar q}$ correspondingly decay as 
\begin{align} \label{e:corrlen1}
\mathcal{C}_{q \bar q} (\bm{B}_1 , k_1^+ \, ; \, \bm{B}_2 , k_2^+) \sim e^{- 2 m |\bm{B}_1 - \bm{B}_2|_T} ,
\end{align}
such that the overall correlation length for $\mathcal{C}_{q \bar q}$ via single-pair production is set by the mass of the quark pair: $L_{q \bar q}^{(single)} = \frac{1}{2 m}$.  If the produced quarks are heavy, such as charm quarks, then the associated correlation length is a perturbative scale.  If the quarks are light (or considered massless), on the other hand, then the correlations can extend to nonperturbatively long distances $|\bm{B}_1 - \bm{B}_2|_T > 1 / \Lambda_{QCD}$.  Our perturbative calculation ceases to be applicable over such long distances, however, so for light quarks one should cut off the correlation function by hand at $|\bm{B}_1 - \bm{B}_2|_T \approx 1 / \Lambda_{QCD}$.

In the heavy-light regime, double-pair production is suppressed relative to single-pair production by a factor of $\alpha_s^2 a^{1/3} \ll 1$, so the quark-antiquark correlation function $\mathcal{C}_{q \bar q}$ is therefore dominated by single-pair production \eqref{e:singlecorr} wherever it contributes.  However, as we saw in \eqref{e:corrlen1}, single-pair production can only accommodate correlations for distances $|\bm{B}_1 - \bm{B}_2|_T \leq \min[ 1/2m \, , \, 1/\Lambda_{QCD}]$; at distances larger than this, the exponential suppression \eqref{e:corrlen1} of the single-pair correlations begins to compete with the suppression by $\alpha_s^2 a^{1/3}$ of the double-pair production mechanism.  Thus for heavy quarks at distances larger than
\begin{align}
r_{double} \equiv \frac{1}{2m} \ln\frac{1}{\alpha_s^2 a^{1/3}} ,
\end{align}
double-pair production becomes the dominant source of $q \bar q$ correlations $\mathcal{C}_{q \bar q}$.  These double-pair mechanisms will have correlation lengths $L_{q \bar q}^{(double)}$ which extend beyond $L_{q \bar q}^{(single)} = 1 / 2 m$.  For $\mathcal{C}_{q q}$ and $\mathcal{C}_{\bar q \bar q}$, single-pair production does not contribute, and double-pair production is the leading mechanism at all length scales.

%
\subsection{The Double-Pair Regime}
%

While we defer a full analysis of the lengthy double-pair production cross-section for a dedicated future publication, there are some general features of double-pair correlations which we can understand already based on the wave function structures \eqref{e:WFpairs} and \eqref{e:WFloop}.  

Consider first the fermion entanglement topology 
\begin{align}
\Psi^{(loop)} &\sim \tr\Big[ \tilde{\Psi} (\bm{x}_1 - \bm{x}_2) 
\tilde{\Psi}^\dagger (\bm{x}_3 - \bm{x}_2) \tilde{\Psi} (\bm{x}_3 - \bm{x}_4)
\tilde{\Psi}^\dagger (\bm{x}_1 - \bm{x}_4) \Big]
\end{align}
shown in Fig.~\ref{f:WFPacMan}.  Any correlation function will tag on the production of quarks or antiquarks at two positions $\bm{B}_1 = \bm{x}_i$ and $\bm{B}_2 = \bm{x}_j$, and the asymptotic limit $r_T = | \bm{B}_1 - \bm{B}_2 |_T \rightarrow \infty$ of the correlation corresponds to sending these coordinates to infinity in opposite directions: $\bm{x}_i = + \half \bm{r}$ and $\bm{x}_j = - \half \bm{r}$.  Because each coordinate $\bm{x}_1 , \bm{x}_2 , \bm{x}_3 , \bm{x}_4$ appears exactly twice in the fermion trace and all wave functions have the same asymptotic decay, all contributions from $\Psi^{(loop)}$ have the long-distance behavior
\begin{align}
\Psi^{(loop)} \sim \left[ e^{- \half m r_T} \right]^4 \sim e^{- 2 m |\bm{B}_1 - \bm{B}_2|_T} .
\end{align}
This exponential suppression is exactly the same as for the single-pair production channel, as one would intuitively expect from a topology in which all the pairs are entangled.  Thus we conclude that, whenever single-pair production is negligible for $\mathcal{C}_{q \bar q}$, so is the fermion entanglement contribution from double-pair production.  For $\mathcal{C}_{q q}$ and $\mathcal{C}_{\bar q \bar q}$, fermion entanglement contributes only at distances shorter than $1/2m$.

Similarly, for the topology with no fermion entanglement shown in Fig.~\ref{f:Sausage_WF}, the wave function structure
\begin{align}
\Psi^{(pairs)} &\sim \tr\Big[ \tilde{\Psi} (\bm{x}_1 - \bm{x}_2) 
\tilde{\Psi}^\dagger (\bm{x}_1 - \bm{x}_2) \Big] 
\tr\Big[\tilde{\Psi} (\bm{x}_3 - \bm{x}_4)
\tilde{\Psi}^\dagger (\bm{x}_3 - \bm{x}_4) \Big]
\end{align}
leads to the same suppression if the size of either pair $|\bm{x}_1 - \bm{x}_2|_T$ or $|\bm{x}_3 - \bm{x}_4|_T$ becomes large.  However, combinations which tag on one (anti)quark from each pair are not suppressed in this way; in these cases, large $|\bm{B}_1 - \bm{B}_2|_T$ requires large separations between the pairs' centers of momentum $|\bm{u}_{12} - \bm{u}_{34}|_T$, rather than that the $q \bar q$ splitting itself be long-distance.  This applies to $\mathcal{C}_{q q}$ and $\mathcal{C}_{\bar q \bar q}$, which necessarily select the quarks or antiquarks from each pair, as well as the specific combinations $\bm{B_1} - \bm{B_2} \in \{ \bm{x_1} - \bm{x_4} , \bm{x_3} - \bm{x_2} \}$ for $\mathcal{C}_{q \bar q}$.  Thus for $\mathcal{C}_{q \bar q}$ at distances $|\bm{B_1} - \bm{B_2}| \gg \frac{1}{m}$, we can replace the delta functions $Z^{(q \bar q)}$ from \eqref{e:Ztags} with just the unsuppressed combinations,
\begin{align}
Z^{(q \bar q)} \rightarrow Z^{(q \bar q)}_{reduced} &= \frac{1}{4} K_1^+ K_2^+ \Big[ \delta^2 (\bm{x_1} - \bm{B_1}) \delta(k_1^+ - K_1^+) \: \delta^2 (\bm{x_4} - \bm{B_2}) \delta(k_4^+ - K_2^+)
\notag \\ & \hspace{1cm} +
\delta^2 (\bm{x_3} - \bm{B_1}) \delta(k_3^+ - K_1^+) \: \delta^2 (\bm{x_2} - \bm{B_2}) \delta(k_2^+ - K_2^+) \Big] .
\end{align}
%

%
\begin{figure}
\includegraphics[width=\textwidth]{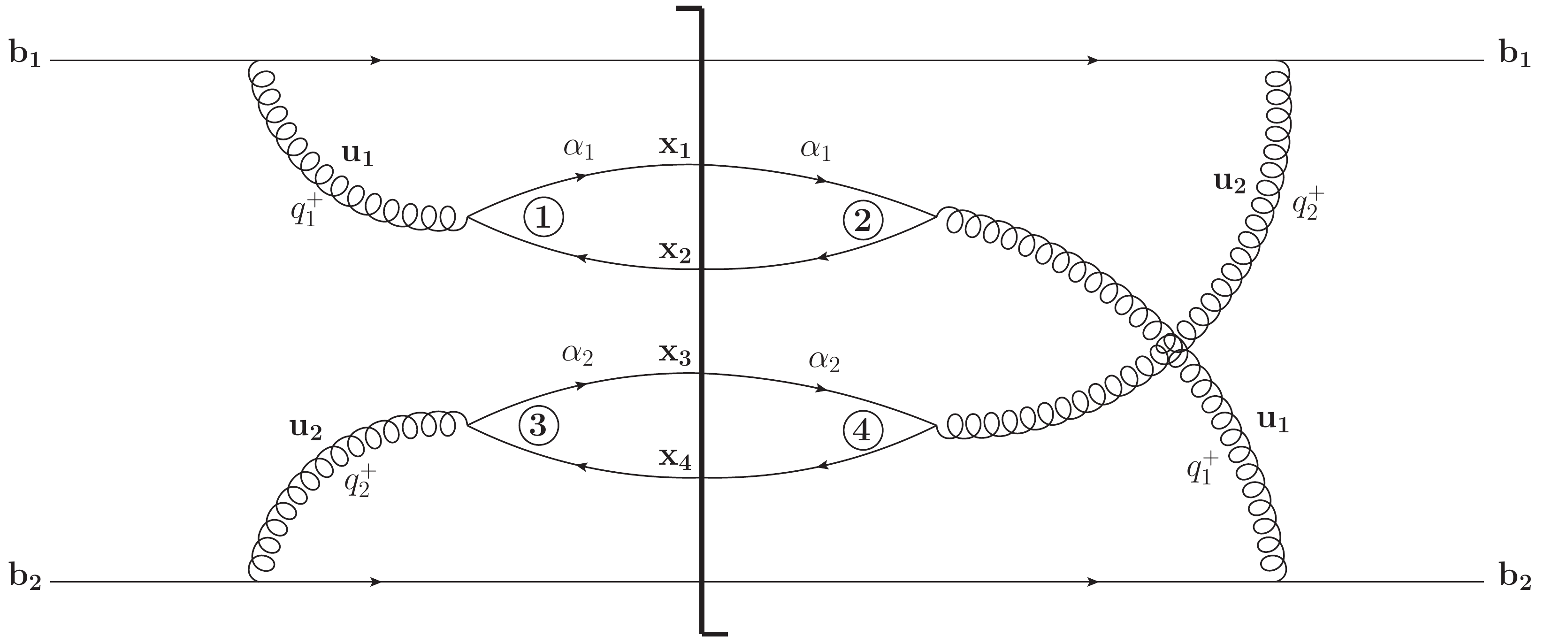}
\caption{ \label{f:Gent} Gluon entanglement channel of double-pair production.} 
\end{figure}
%

For heavy quarks, the region $r_{double} \ll |\bm{B}_1 - \bm{B}_2|_T \ll \frac{1}{\Lambda_{QCD}}$ is especially interesting.  Here the distance is large enough that the dominant production process is double-pair production (with no fermion entanglement), but it is also short enough that correlations can arise from perturbative QCD.  These correlations can occur either through the interactions of the pairs with the same correlated color domains in the heavy nucleus, or through the entanglement of the gluons which produce the two pairs, as illustrated in Fig.~\ref{f:Gent}.

For nonperturbatively large distances $|\bm{B_1} - \bm{B_2} |_T \gsim \frac{1}{\Lambda_{QCD}}$, perturbative degrees of freedom cannot mediate the correlations.  Entanglement of gluons, as in Fig.~\ref{f:Gent}, is not possible in this region because the range of perturbative gluons is limited to $1/\Lambda_{QCD}$.  In this case, the only contribution to the double-pair cross-section \eqref{e:CSdouble} occurs when the large separation in $|\bm{B}_1 - \bm{B}_2|_T$ is due to a large separation between the nucleons $|\bm{b}_1 - \bm{b}_2|_T$ which radiate the independent pairs.  At these distances, the two pairs are truly independent: they are not connected by the exchange of any perturbative degrees of freedom, and they interact with disjoint, uncorrelated color fields in the heavy nucleus.  Because of this, both the wave functions and the interactions of the two pairs factorize; for example, for $\mathcal{C}_{q q}$ we have
%
\begin{align} \label{e:geomregion1}
K_1^+ K_2^+ \frac{d\sigma^{(q q)}}{d^2 B_1 dK_1^+ \, d^2 B_2 dK_2^+} &= 
\int d^2 B \: \int\left( \prod_{i=1}^4 d^2 x_i \frac{dk_i^+}{4\pi k_i^+} \right)
Z^{(q q)} (\bm{B}_1 , K_1^+ , \bm{B}_2 , K_2^+ , \{ \bm{x}_i , k_i^+ \}_{i=1}^4)
\notag \\ &\times
\left[ \int d^2 b_1 \, T_a (\bm{b_1} - \bm{B}) 
\left\langle | \tilde{\mathcal{A}}_{N A} (\bm{b}_1 ; \bm{x}_1 , k_1^+ , \bm{x}_2 , k_2^+) |^2 \right\rangle
\right] 
\notag \\ &\times
\left[ \int d^2 b_2 \, T_a (\bm{b_2} - \bm{B}) 
\left\langle | \tilde{\mathcal{A}}_{N A} (\bm{b}_2 ; \bm{x}_3 , k_3^+ , \bm{x}_4 , k_4^+) |^2 \right\rangle
\right]   
\notag \\ \notag  \\ &=
\int d^2 B 
\left( K_1^+ \frac{d\sigma^{q}}{d^2 B_1 dK_1^+ \: d^2 B} \right) \:
\left( K_2^+ \frac{d\sigma^{q}}{d^2 B_2 dK_2^+ \: d^2 B} \right) ,
\end{align}
where we integrated out the spectators in the last line.  Still, even in this regime the correlated part \eqref{e:geomregion1} does not fully factorize into the product of two uncorrelated cross-sections, because the two pairs are constrained to arise from nucleons in the same light nucleus $T_a(\bm{b})$.  These are the ``geometric correlations'' observed in \cite{Kovchegov:2012nd}, which contribute at distances $|\bm{B}_1 - \bm{B}_2|_T$ up to the size $R_a$ of the light nucleus.  

In the absence of geometrical structure from the light ion, say 
\begin{align}
K_1^+ \frac{d\sigma^q}{d^2 B_1 dK_1^+ \: d^2 B} = \frac{1}{S_{\bot \, A}} \: 
\left(K_1^+ \frac{d\sigma^q}{d^2 B_1 dK_1^+} \right) ,
\end{align}
with $S_{\bot \, A}$ the transverse area of the heavy nucleus, a full factorization of the correlated part \eqref{e:geomregion1} does occur, such that the overall correlation function reduces to
\begin{align}
\mathcal{C}_{q q} (\bm{B}_1 , K_1^+ \, ; \, \bm{B}_2 , K_2^+) = \left[ \frac{\sigma_{inel}}{S_{\bot \, A}} - 1 \right]
\left\langle K_1^+ \frac{dn^q}{d^2 B_1 K_1^+} \right\rangle
\left\langle K_2^+ \frac{dn^q}{d^2 B_2 K_2^+} \right\rangle.
\end{align}
Even though there are no correlations of any kind, the correlation function itself remains nonzero, and the situation is similar for $\mathcal{C}_{q \bar q}$ and $\mathcal{C}_{\bar q \bar q}$.  This is a consequence of the simple definition \eqref{e:corr1}; other definitions \cite{Kovchegov:2012nd} can be chosen such that the correlations go to zero here instead.

%
\section{Results and Outlook}
\label{sec:concl} 
%

%
\subsection{Results: Single-Pair Production}
%

As a concrete illustration, let us directly compute the single-pair contribution to the quark-antiquark correlation function $\mathcal{C}_{q \bar q}$ \eqref{e:qaqcorr1}.  The crux is the evaluation of the cross-section \eqref{e:singlecorr} which gives the correlated part of $\mathcal{C}_{q \bar q}$, for which we will employ the MV / GBW model \eqref{e:MV}.  For simplicity, we will take the large-$N_c$ limit, in which
\begin{align}
\left\langle \Dtwo{x}{y} \Dtwo{u}{v} \right\rangle \approx \left\langle \Dtwo{x}{y} \right\rangle \,
\left\langle \Dtwo{u}{v} \right\rangle = e^{-\frac{1}{4} |\bm{x} - \bm{y}|_T^2 Q_s^2} \:\:
e^{-\frac{1}{4} |\bm{u} - \bm{v}|_T^2 Q_s^2} .
\end{align}
Organizing the terms based on the interactions, the correlated part is
\newpage
\begin{align}
k_1^+ k_2^+ & \frac{d\sigma^{q \bar q}}{d^2 B_1 dk_1^+ \, d^2 B_2 dk_2^+} = 
\frac{a}{(4\pi)^2} \frac{N_c}{2} \int d^2 b_1 
\notag \\ & 
\times \Bigg\{ \Big( \mathcal{U}_1 \mathcal{U}_1 - \mathcal{L}_1 \mathcal{L}_1 + 
\bm{\mathcal T}_1 \cdot \bm{\mathcal T}_1 + \mathcal{U}_1 \mathcal{U}_2 - 
\mathcal{L}_1 \mathcal{L}_2 + \bm{\mathcal T}_1 \cdot \bm{\mathcal T}_2\Big) 
(\bm{B}_1 - \bm{B}_2 , \bm{u} - \bm{b}_1 , \alpha)
\notag \\ & \hspace{1cm}
\times \Bigg[ 1 - e^{-\frac{1}{4} \left[ |\bm{B}_1 - \bm{b}_1 |_T^2 + |\bm{B}_2 - \bm{b}_1 |_T^2 \right] Q_s^2} \Bigg]
\notag \\ &  
+ \Big( \mathcal{U}_2 \mathcal{U}_2 - \mathcal{L}_2 \mathcal{L}_2 + 
\bm{\mathcal T}_2 \cdot \bm{\mathcal T}_2 + \mathcal{U}_1 \mathcal{U}_2 - 
\mathcal{L}_1 \mathcal{L}_2 + \bm{\mathcal T}_1 \cdot \bm{\mathcal T}_2\Big) 
(\bm{B}_1 - \bm{B}_2 , \bm{u} - \bm{b}_1 , \alpha)
\notag \\ & \hspace{1cm}
\times \Bigg[ 1 - e^{-\frac{1}{2} |\bm{u} - \bm{b}_1|_T^2 Q_s^2} \Bigg] 
\notag \\ & 
- \Big( \mathcal{U}_1 \mathcal{U}_2 - \mathcal{L}_1 \mathcal{L}_2 + 
\bm{\mathcal T}_1 \cdot \bm{\mathcal T}_2 \Big) (\bm{B}_1 - \bm{B}_2 , \bm{u} - \bm{b}_1 , \alpha)
\notag \\ & \hspace{1cm}
\times \Bigg[ 1 - e^{-\frac{1}{4} \left[ |\bm{B}_1 - \bm{u}|_T^2 + |\bm{B}_2 - \bm{u}|_T^2 \right] Q_s^2} \Bigg]
\Bigg\} ,
\end{align}
where we recall that $\bm{u} = \alpha \bm{B}_1 + (1-\alpha) \bm{B}_2$ is the center of momentum of the $q \bar q$ pair and $\alpha = \frac{k_1^+}{k_1^+ + k_2^+}$ is the longitudinal momentum fraction of the pair carried by the quark.  We note that the saturation scale $Q_s$ contains an impact parameter dependence which can in principle be different from term to term.  However, $Q_s^2 (\bm{b}) \propto T_A(\bm{b})$ varies only over macroscopic scales proportional to $A^{1/3}$, over which the perturbatively short-distance changes in coordinates are negligible; we can therefore choose to evaluate $Q_s$ at the same position (say, $\bm{u}$) in all terms up to corrections of order $\ord{A^{-1/3}}$.  Changing integration variables from $\bm{b}_1$ to $\bm{w} = \bm{u} - \bm{b}_1$ and employing the short-hand $\bm{r} = \bm{B}_1 - \bm{B}_2$ gives
\begin{align} \label{e:pair1}
k_1^+ k_2^+ & \frac{d\sigma^{q \bar q}}{d^2 B_1 dk_1^+ \, d^2 B_2 dk_2^+} = 
\frac{a}{(4\pi)^2} \frac{N_c}{2} \int d^2 w 
\notag \\ & 
\times \Bigg\{ \Big( \mathcal{U}_1 \mathcal{U}_1 - \mathcal{L}_1 \mathcal{L}_1 + 
\bm{\mathcal T}_1 \cdot \bm{\mathcal T}_1 + \mathcal{U}_1 \mathcal{U}_2 - 
\mathcal{L}_1 \mathcal{L}_2 + \bm{\mathcal T}_1 \cdot \bm{\mathcal T}_2\Big) 
(\bm{r} , \bm{w} , \alpha)
\notag \\ & \hspace{1cm}
\times \Bigg[ 1 - e^{-\half w_T^2 Q_s^2} \, e^{-\frac{1}{4} \left[ \alpha^2 + (1-\alpha)^2 \right] r_T^2 Q_s^2} \,
e^{- \half (1-2\alpha) \cos\phi \, w_T r_T Q^2} \Bigg]
\notag \\ &  
+ \Big( \mathcal{U}_2 \mathcal{U}_2 - \mathcal{L}_2 \mathcal{L}_2 + 
\bm{\mathcal T}_2 \cdot \bm{\mathcal T}_2 + \mathcal{U}_1 \mathcal{U}_2 - 
\mathcal{L}_1 \mathcal{L}_2 + \bm{\mathcal T}_1 \cdot \bm{\mathcal T}_2\Big) 
(\bm{r} , \bm{w} , \alpha)
\: \Bigg[ 1 - e^{-\frac{1}{2} w_T^2 Q_s^2} \Bigg] 
\notag \\ & 
- \Big( \mathcal{U}_1 \mathcal{U}_2 - \mathcal{L}_1 \mathcal{L}_2 + 
\bm{\mathcal T}_1 \cdot \bm{\mathcal T}_2 \Big) (\bm{r} , \bm{w} , \alpha)
\: \Bigg[ 1 - e^{-\frac{1}{4} \left[ \alpha^2 + (1-\alpha)^2 \right] r_T^2 Q_s^2} \Bigg] \Bigg\} .
\end{align}
The exponentials in the first set of terms above come from the interference of $\Psi_1$ and $\Psi_3$ from Fig.~\ref{f:WF}, the exponentials in the second set of terms comes from the interference of $\Psi_2$ and $\Psi_3$, and the exponentials in the last set of terms comes from the interference of $\Psi_1$ and $\Psi_2$.  The terms with $1$ in brackets come from the squares $|\Psi_1|^2$, $|\Psi_2|^2$, and $|\Psi_3|^2$.

The $d^2 w$ integral is involved but straightforward, and some of the terms contain a logarithmic IR divergence reflecting the (infinite) total radiation produced by the bare valence quark.  Keeping the terms which are logarithmic in the IR cutoff $\Lambda$ and dropping the finite pieces, the result is
\begin{align} \label{e:stuff}
k_1^+ k_2^+ \frac{d\sigma^{q \bar q}}{d^2 B_1 dk_1^+ \, d^2 B_2 dk_2^+} &= 
\left( \frac{a \, \alpha_s^2 N_c}{4\pi^3} \ln\frac{1}{\Lambda} \right) \alpha (1-\alpha) m^2 \,
\left( 1 - e^{-\frac{1}{4} [\alpha^2 + (1-\alpha)^2] r_T^2 Q_s^2} \right)
\notag \\ &\times
\left[ \left( \alpha^2 + (1-\alpha)^2 \right) K_1^2 (m r_T) + K_0^2 (m r_T) \right].
\end{align}
Note that we have not specified the scale in numerator of the logarithm, since choices of that scale differ only by finite pieces.  In the same approximations, the inelastic cross-section \eqref{e:siginel} becomes
\begin{align} 
\sigma_{inel} &= \frac{\alpha_s N_c}{\pi^2} \, (a \, \Delta Y) \, \int \frac{ d^2 x_1 \, d^2 x_2}{x_{21}^2}
\left( 1 - e^{-\half x_{21}^2 Q_s^2} \right)
\notag \\ &=
\frac{2 \alpha_s N_c}{\pi} \, (a \, S_{\bot \, A} \, \Delta Y) \ln\frac{1}{\Lambda} ,
\end{align}
with $S_{\bot \, A} \propto A^{2/3}$ the transverse area of the heavy target nucleus.  The IR logarithms cancel in the ratio, leaving the correlated part of $\mathcal{C}_{q \bar q}$ as
\begin{align} \label{e:corrpart}
\mathcal{C}_{q \bar q}^{corr} (\bm{B}_1 , k_1^+ ; \bm{B}_2 , k_2^+) &\equiv \frac{1}{\sigma_{inel}} 
\left( k_1^+ k_2^+ \frac{d\sigma^{q \bar q}}{d^2 B_1 dk_1^+ \, d^2 B_2 dk_2^+} \right)
\notag \\ &=
\frac{\alpha_s}{8 \pi^2} \left( \frac{1}{S_{\bot \, A} \, \Delta Y} \right) \alpha (1-\alpha) m^2 \,
\left( 1 - e^{-\frac{1}{4} [\alpha^2 + (1-\alpha)^2] r_T^2 Q_s^2} \right)
\notag \\ &\times
\left[ \left( \alpha^2 + (1-\alpha)^2 \right) K_1^2 (m r_T) + K_0^2 (m r_T) \right].
\end{align}

We observe that the fully differential correlation function $\mathcal{C}_{q \bar q}^{corr} (\bm{B}_1 , k_1^+ ; \bm{B}_2 , k_2^+)$ is suppressed by a factor of the full phase space $\frac{1}{S_{\bot \, A} \, \Delta Y}$.  A more meaningful correlation function is therefore one in which the center-of-mass degrees of freedom $\bm{u}$ and $q^+ = k_1^+ + k_2^+$ have been integrated out to cancel this factor.  Before we do that, however, let us first integrate out the antiquark degrees of freedom $\bm{B}_2 , k_2^+$ to obtain the inclusive quark production cross-section which contributes to the uncorrelated part of the correlation function $\mathcal{C}_{q \bar q}$:
\begin{align}
\frac{1}{\sigma_{inel}} \left( k_1^+ \frac{d\sigma^{q}}{d^2 B_1 dk_1^+} \right) &=
\int\limits_{p_A^+}^{p_a^+} \frac{dk_2^+}{k_2^+} \int d^2 B_2 \, 
\mathcal{C}_{q \bar q}^{corr} (\bm{B}_1 , k_1^+ ; \bm{B}_2 , k_2^+) .
\end{align}
The transverse integral can be done analytically in terms of Meijer $G$ functions, but it will not generate a factor of the area $S_{\bot \, A}$ because the Bessel functions $K_1^2 (m r_T) , K_0^2 (m r_T)$ limit the size of the splitting to be of order $\ord{1/m}$.  As a consequence, the inclusive quark production term is suppressed by a factor of the transverse area,
\begin{align}
\frac{1}{\sigma_{inel}} \left( k_1^+ \frac{d\sigma^{q}}{d^2 B_1 dk_1^+} \right) \propto \frac{1}{S_{\bot \, A}}
\propto A^{-2/3}.
\end{align}
Therefore, in the full correlation function \eqref{e:qaqcorr1}, the correlated part $\mathcal{C}^{corr}_{q \bar q}$ is proportional to $1 / S_{\bot \, A}$, while the uncorrelated part $(\sigma^q / \sigma_{inel})  (\sigma^{\bar q} / \sigma_{inel})$ is proportional to $(1 / S_{\bot \, A})^2$.  The uncorrelated part is thus suppressed by a factor of $A^{-2/3}$ relative to the correlated part and can be neglected: $\mathcal{C}_{q \bar q} \approx \mathcal{C}_{q \bar q}^{corr}$.

After integrating out the center-of-mass degrees of freedom $\bm{u}$ and $q^+$ for fixed $\bm{r}$ and $\alpha$, the correlation function is
\begin{align}
\mathcal{C}_{q \bar q} (\bm{r} , \alpha) &\equiv \frac{1}{\sigma_{inel}} \frac{d\sigma^{q \bar q}}{d^2 r \, d\alpha}
= \int\limits_{p_A^+}^{p_a^+} \frac{dq^+}{q^+} \int d^2 u \, \frac{1}{\sigma_{inel}} 
\left( q^+ \frac{d\sigma^{q \bar q}}{d^2 r \, d\alpha \, d^2 u \, dq^+} \right)
\notag \\ &=
\frac{1}{\alpha (1-\alpha)} \int\limits_{p_A^+}^{p_a^+} \frac{dq^+}{q^+} \int d^2 u \, \frac{1}{\sigma_{inel}} 
\left( k_1^+ k_2^+ \frac{d\sigma^{q \bar q}}{d^2 B_1 \, dk_1^+ \, d^2 B_2 \, dk_2^+} \right) .
\end{align}
The $dq^+$ integration of \eqref{e:corrpart} trivially cancels the rapidity phase space factor $\Delta Y$, and the $d^2 u$ integration only couples to the impact parameter dependence of the saturation scale:
\begin{align}
\int \frac{d^2 u}{S_{\bot \, A}} \left( 1 - e^{-\frac{1}{4} [ \alpha^2 + (1-\alpha)^2 ] r_T^2 Q_s^2 (\bm{u})} \right) =
1 - \int \frac{d^2 u}{S_{\bot \, A}} \, e^{-\frac{1}{4} [ \alpha^2 + (1-\alpha)^2 ] r_T^2 Q_s^2 (\bm{u})} .
\end{align}
Since this impact parameter dependence corresponds to the nuclear profile function $Q_s^2 (\bm{u}) \propto T_A (\bm{u})$, the precise result of this averaging depends on the geometry of the target nucleus.  One can explore the effects of different nuclear geometries like a Woods-Saxon profile, but for our present purposes we will simply neglect the impact parameter dependence of $Q_s$, effectively treating the target nucleus as having uniform transverse density.  In this approximation, the averaging is trivial and cancels the area factor $S_{\bot \, A}$.

Having cancelled the phase space factor $\frac{1}{S_{\bot \, A} \Delta Y}$, the resulting correlation function is
\begin{align}
\mathcal{C}_{q \bar q} (\bm{r} , \alpha) = \frac{\alpha_s}{8 \pi^2} m^2 
\left(1 - e^{-\frac{1}{4} [ \alpha^2 + (1-\alpha)^2] r_T^2 Q_s^2} \right) 
\left[ \left( \alpha^2 + (1-\alpha)^2 \right) K_1^2 (m r_T) + K_0^2 (m r_T) \right] .
\end{align}
This correlation function contains three-dimensional information about the distribution of quarks and antiquarks generated through single pair production, but since the splitting fraction $\alpha$ is not fixed in a given event, it may be desirable to integrate out $\alpha$ as well.  Doing so gives the final result for the $q \bar q$ correlation function,
\begin{align} \label{e:2Dcorr}
\mathcal{C}_{q \bar q} (\bm{r}) &\equiv \int_0^1 d\alpha \, \mathcal{C}_{q \bar q} (\bm{r}, \alpha)
\notag \\ &=
\frac{\alpha_s}{8 \pi^2} m^2 \Bigg\{ K_1^2 (m r_T) \Bigg[ \frac{2}{3} + \frac{2}{r_T^2 Q_s^2} \,
e^{-\frac{1}{4} r_T^2 Q_s^2} - \sqrt{\frac{\pi}{2}} \, \frac{4+r_T^2 Q_s^2}{r_T^3 Q_s^3} \, e^{-\frac{1}{8} r_T^2 Q_s^2} \, \mathrm{Erf} \left( \frac{r_T Q_s}{2 \sqrt{2}} \right) \Bigg] 
\notag \\ & \hspace{2cm} +
K_0^2 (m r_T) \Bigg[ 1 - \frac{\sqrt{2 \pi}}{r_T Q_s} \, e^{-\frac{1}{8} r_T^2 Q_s^2} \, 
\mathrm{Erf} \left( \frac{r_T Q_s}{2 \sqrt{2}} \right) \Bigg] \Bigg\} .
\end{align}
At this accuracy, the double-pair production cross-section is suppressed and can be neglected, so that the baryon number correlation function \eqref{e:Bcorr1} is given entirely by the $q \bar q$ correlation functions:
\begin{align} \label{e:2Dbaryon}
\mathcal{C}_{\mathcal{B} \mathcal{B}} (\bm{r}) = -\frac{1}{9} \sum_f \Bigg( \mathcal{C}_{q_f \bar{q}_f} (\bm{r}) +
\mathcal{C}_{q_f \bar{q}_f} (-\bm{r}) \Bigg) .
\end{align}
The expressions \eqref{e:2Dcorr} and \eqref{e:2Dbaryon} are one of the primary results of this paper, and they illustrate what we hope to achieve in future work which will extend this result to include correlations from double-pair production.  We note that, although much of the physical content of this correlation is due to the mass of the quarks, Eq.~\eqref{e:2Dcorr} does have a well-behaved massless limit, for which $\lim_{m \rightarrow 0} [ m^2 K_1^2 (m r_T)]  = \frac{1}{r_T^2}$ and $\lim_{m \rightarrow 0} [ m^2 K_0^2 (m r_T) ] = 0$.

%
\begin{figure}
\begin{center}
\includegraphics[width=1\textwidth]{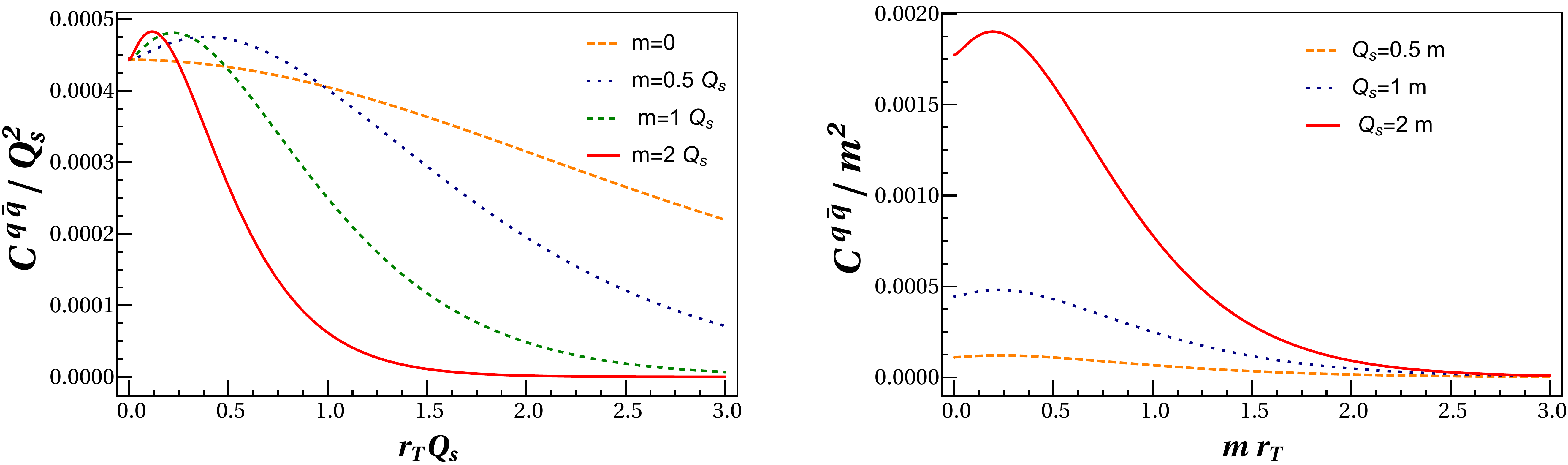}
\end{center}
\vspace{-1cm}
\caption{Functional dependence of the correlation function \eqref{e:2Dcorr} on the quark mass (left panel) and saturation scale (right panel).} 
\label{f:TheoryPlots}
\end{figure}
%

%
\begin{figure}
\begin{center}
\includegraphics[width=1\textwidth]{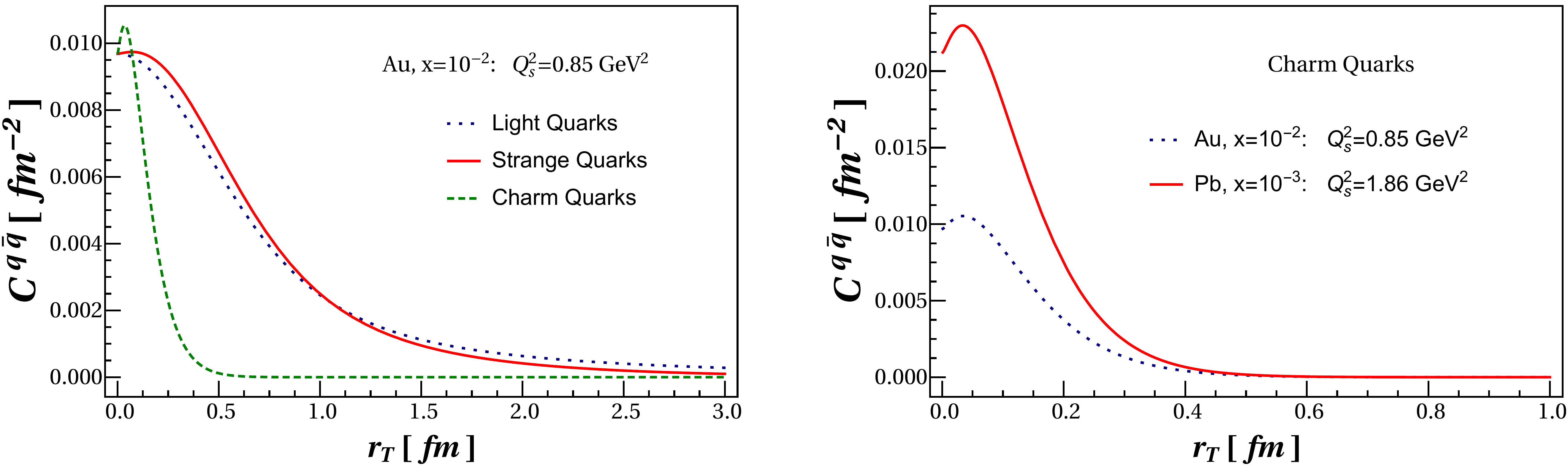}
\end{center}
\vspace{-1cm}
\caption{Evaluation of the correlation function \eqref{e:2Dcorr} for physical quark masses and with saturation scales chosen to reflect collisions with gold ions at RHIC and lead ions at the LHC.} 
\label{f:ExptPlots}
\end{figure}
%

%
\begin{figure}
\begin{center}
\includegraphics[width=0.5\textwidth]{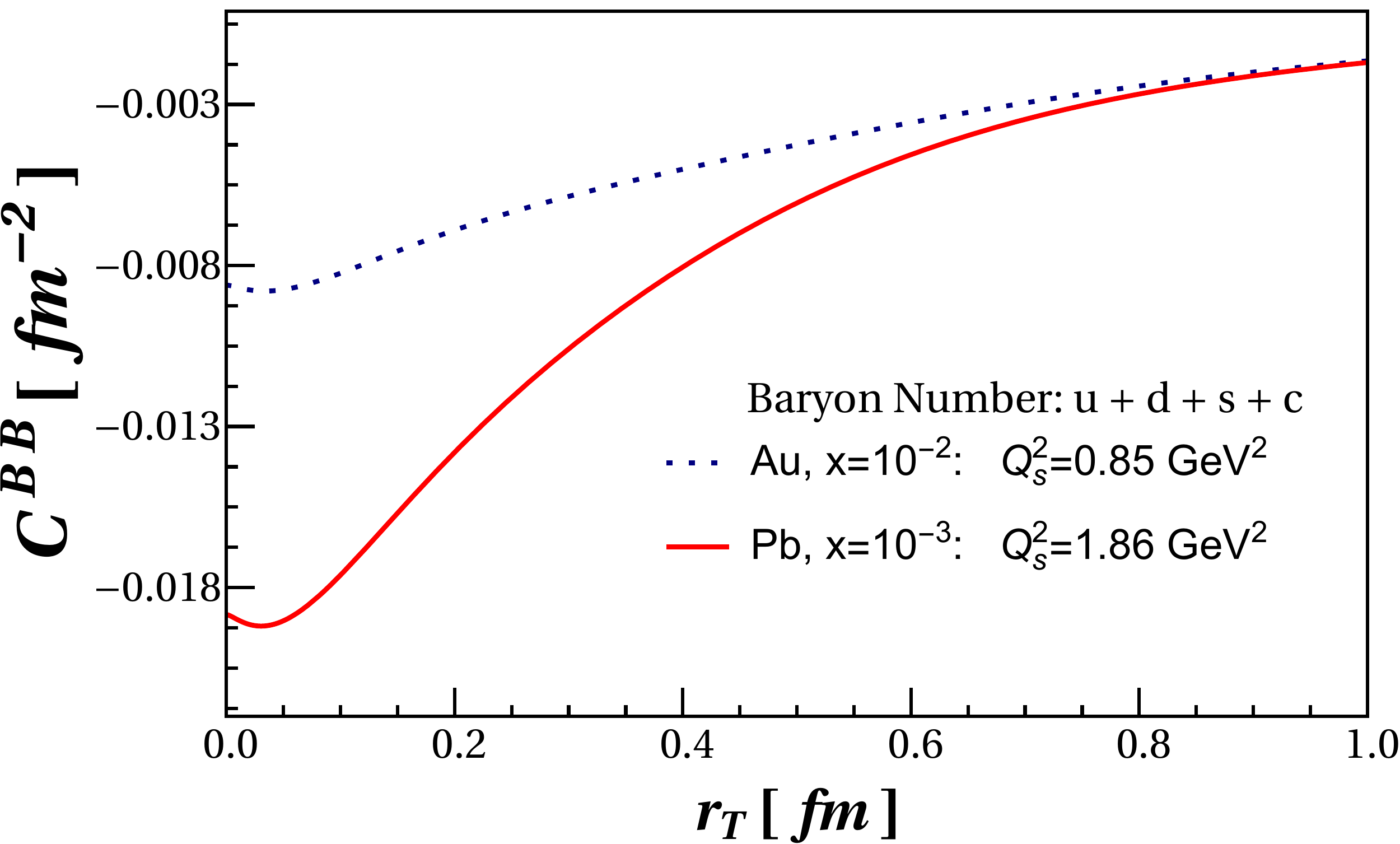}
\end{center}
\vspace{-1cm}
\caption{Evaluation of the baryon number correlation function \eqref{e:2Dbaryon} for physical quark masses, summing over up, down, strange, and charm quark flavors.} 
\label{f:BaryonPlot}
\end{figure}
%

The functional dependence of \eqref{e:2Dcorr} on $m$ and $Q_s$ is illustrated in Fig.~\ref{f:TheoryPlots}, and the correlation function for physical values of the parameters is shown in Fig.~\ref{f:ExptPlots}.  In these plots, we take $\alpha_s = 0.3$ for our fixed-coupling calculation, and we choose the saturation scales to reflect gold ions at $x = 10^{-2}$ at RHIC and lead ions at $x = 10^{-3}$ at the LHC.  For the former case, we take the value of $Q_s$ quoted in the IPsat model (see, for instance, \cite{Mantysaari:2017slo}), and for the latter we extrapolate using the rough pocket formula $Q_s^2 \propto (\frac{A}{x})^{1/3}$.  As expected, the range of the single-pair $q \bar q$ correlation function \eqref{e:2Dcorr} is controlled by the quark mass and not by the saturation scale.  However, as seen in Fig~\ref{f:TheoryPlots}, the saturation scale does control the strength of the correlations.  The corresponding plot for the baryon number correlation function \eqref{e:2Dbaryon} is shown in Fig.~\ref{f:BaryonPlot}.  Here we have summed over up, down, strange, and charm quark flavors; note that the negative value of $\mathcal{C}_{\mathcal{B} \mathcal{B}}$ (anticorrelation) reflects the conditional probability of finding a negative baryon number charge in the vicinity of an associated positive one (and vice versa). 

It is also interesting to examine the diagrammatic origins of the correlation function \eqref{e:2Dcorr}. The terms containing the IR logarithms which generated \eqref{e:2Dcorr} all arose from the last line of Eq.~\eqref{e:pair1}, which received contributions from interference of the time orderings $\Psi_1$ and $\Psi_2$ (the first and second diagrams of Fig.~\ref{f:buildblock}) as well as the square $|\Psi_3|^2= | - \Psi_1 - \Psi_2 |^2$ (the third diagram in Fig.~\ref{f:buildblock}). As such, the dynamical part of the correlation function \eqref{e:2Dcorr} reflects the correlations generated by the interference of the $q\bar{q}$ pair scattering and the gluon scattering: the stronger the effects of multiple scattering (quantified by $Q_s$), the stronger the resulting correlations. These correlations are therefore genuine effects of multiple scattering which would would be absent or weak in minimum-bias pp collisions, but enhanced in pA collisions. They also dominate the correlations present over short distances in heavy-light ion collisions, although as discussed previously, at longer distances double-pair production mechanisms will begin to dominate. 

We emphasize the crucial role of the time ordering $\Psi_2$ in generating the quark/antiquark correlation function \eqref{e:2Dcorr}.  This time ordering, in which the radiated gluon first scatters in the target field and then pair produces in the final state, is responsible for both the exponential dipole term in Eq.~\eqref{e:stuff} by its interference with the time ordering $\Psi_1$ and for the 1 by its contribution to $|\Psi_3|^2= | - \Psi_1 - \Psi_2 |^2$.  It is precisely the contribution $\Psi_2$ which was omitted in the Pauli blocking calculation of~\cite{Altinoluk:2016vax} because it was not expected to generate statistical quark/quark correlations.  In our case, however, the $\Psi_2$ contribution is essential to obtaining the interaction-mediated quark/antiquark correlations \eqref{e:2Dcorr}, where statistical effects are absent.  

Last, we note that the approximations used to arrive at our results are rather simple ones – the MV / GBW model for the interactions, the large-$N_c$ limit, and the back-of-the-envelope setting of $Q_s$. They illustrate the physical picture of the correlations and provide concrete benchmarks, but these approximations can and should be improved in future work.

%
\subsection{The Physical Picture}
%

%
\begin{figure}
\includegraphics[width=\textwidth]{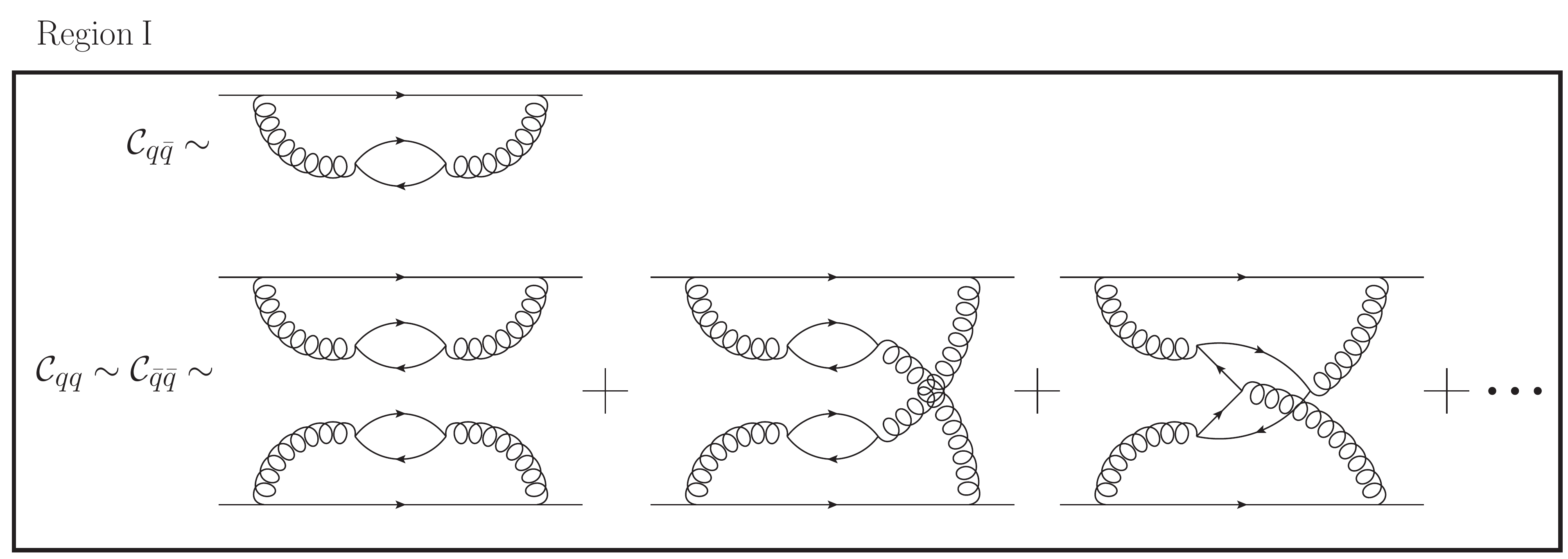}
\includegraphics[width=\textwidth]{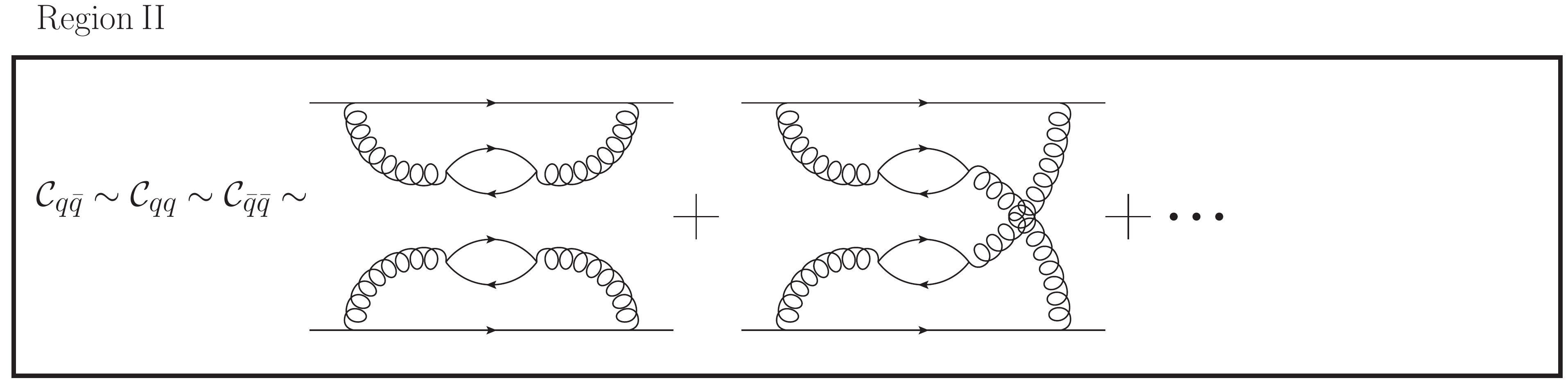}
\includegraphics[width=\textwidth]{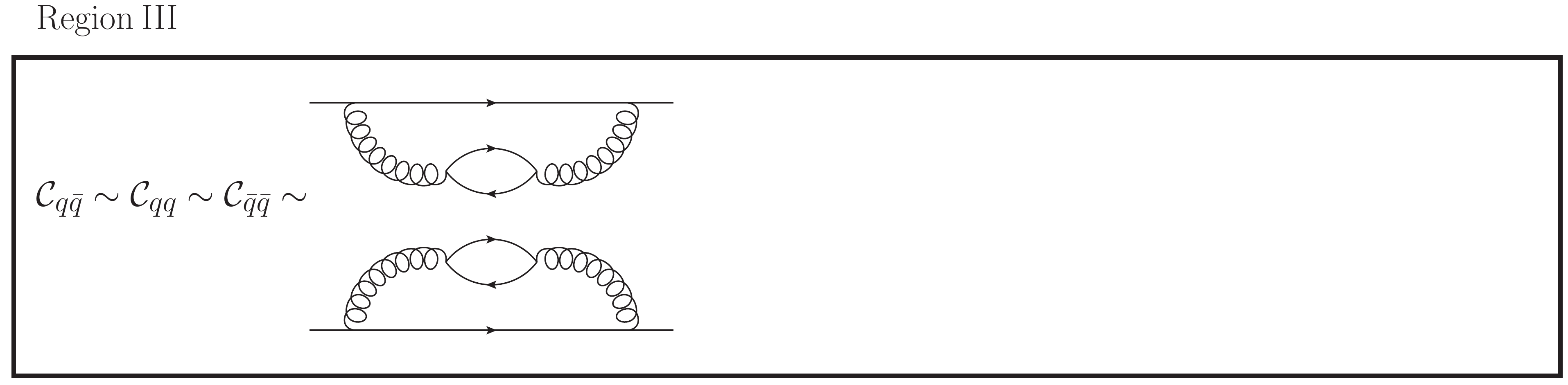}
\caption{Example diagrams contributing to $\mathcal{C}_{q q}$ , $\mathcal{C}_{q \bar q}$ , and $\mathcal{C}_{\bar q \bar q}$ in Regions I, II, and III.} 
\label{f:R1}
\end{figure}
%

As a result of the previous analysis, we can identify three different regimes (see Fig.~\ref{f:R1}) in which the correlation functions $\mathcal{C}_{q q} , \mathcal{C}_{q \bar q} , \mathcal{C}_{\bar q \bar q}$ will be dominated by different mechanisms:
\begin{subequations}
\begin{align}
\mathrm{Region \: I:} & \hspace{1cm} 
|\bm{B_1} - \bm{B_2}|_T \leq \min\left[ \frac{1}{2m} \ln\frac{1}{\alpha_s^2 a^{1/3}} \, , \, \frac{1}{\Lambda_{QCD}} \right]
\\
\mathrm{Region \: II:} & \hspace{1cm} 
\frac{1}{2m} \ln\frac{1}{\alpha_s^2 a^{1/3}} \ll |\bm{B_1} - \bm{B_2}|_T \ll \frac{1}{\Lambda_{QCD}}
\\
\mathrm{Region \: III:} & \hspace{1cm} 
\frac{1}{\Lambda_{QCD}} \leq |\bm{B_1} - \bm{B_2}|_T \leq R_a .
\end{align}
\end{subequations}

Region I covers distances short enough that a single $q \bar q$ splitting can generate the correlations.  For $\mathcal{C}_{q \bar q}$, the dominant mechanism is single-pair production, and the relatively simple result is given in Eqs.~\eqref{e:singlecorr} and \eqref{e:2Dcorr}.  For $\mathcal{C}_{q q}$ and $\mathcal{C}_{\bar q \bar q}$, the full complexity of double-pair production contributes, including fermion entanglement and gluon entanglement.  This maximally complex situation couples to high-order color multipoles, like sextupoles and octupoles, which are difficult to handle analytically, so it may be necessary to resort to numerical~\cite{Dumitru:2011vk} or approximate methods like the large-$N_c$ limit~\cite{Dominguez:2012ad,Iancu:2011ns,Iancu:2011nj}.

Region II only exists for heavy quarks; it covers distances large enough that single-pair and fermion-entanglement contributions have died off, but small enough that perturbative mechanisms still generate the correlations.  All correlations are dominated by double-pair production without fermion entanglement, although perturbative correlations are generated by gluon entanglement and the interactions.  The interactions in this case, while simpler and fewer in number than when fermion entanglement is considered, still couple to high-order color multipoles and may require numerical simulations or approximate analytical methods such as large $N_c$ limit.

Region III covers the longest distances over which correlations exist.  Over these nonperturbative scales, two independent pairs are produced without dynamic correlations from entanglement or the interactions.  Despite this, nontrivial correlations still persist due to the geometric correlations of the pairs from being produced within the density profile of the light nucleus.

\begin{figure}
\includegraphics[width=\textwidth]{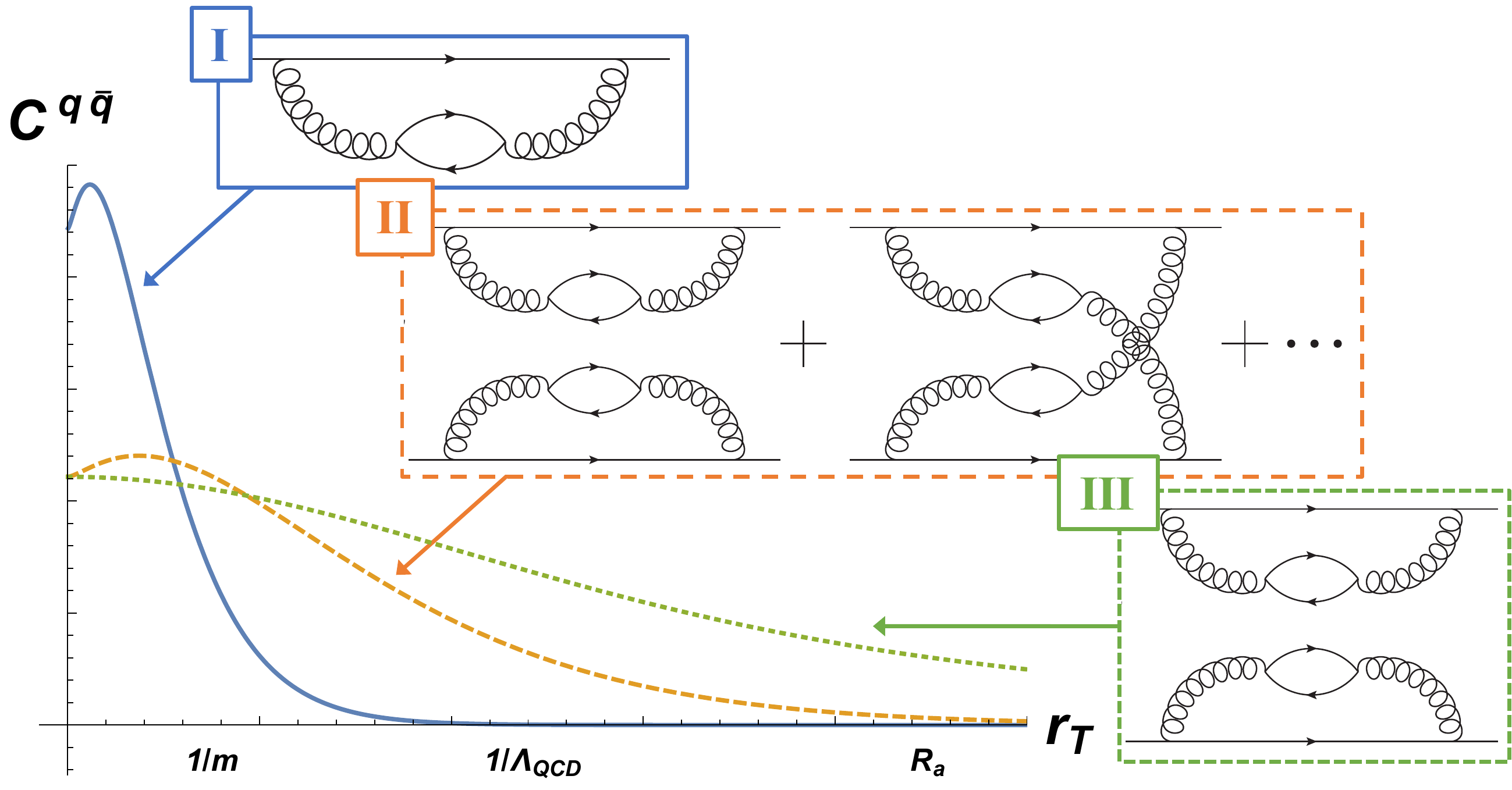}
\caption{Cartoon illustrating the general expected form of the correlation function $\mathcal{C}^{q \bar q}$.  The single-pair production channel calculated in \eqref{e:2Dcorr} dominates in Region I, with a typical range of $1/m$.  The double-pair channels dominating Regions II and III are drawn schematically here; compared to the single-pair correlations in Region I, these channels are suppressed in magnitude by a factor of $\alpha_s^2 a^{1/3}$ but persist over longer ranges.  The interaction- and entanglement-mediated correlations which characterize Region II are cut off at $\ord{\frac{1}{\Lambda_{QCD}}}$, while geometric correlations persist in Region III up to the size $R_a$ of the light ion.} 
\label{f:cartoon}
\end{figure}

This overall physical picture is presented in Fig.~\ref{f:cartoon} for the case of $\mathcal{C}_{q \bar q}$; this cartoon illustrates the transition from the single-pair mechanism calculated in \eqref{e:2Dcorr} for Region I to the double-pair mechanisms in Regions II and III.  The physical picture presented in Fig.~\ref{f:cartoon}, the associated diagrams presented in Fig.~\ref{f:R1}, and the ingredients with which to calculate them -- the wave functions in \eqref{e:WFpairs} and \eqref{e:WFloop} and the Wilson lines in \eqref{e:Smatrix1} -- are the second main result of this paper.  With these components, constructing the double-pair correlation function is straightforward but lengthy, as the particular example \eqref{e:intexample} illustrates.  We will defer a detailed analysis of the double-pair results for a dedicated paper in Part II of this study.

%
%

%
\subsection{Conclusions}
%

In this paper, we have constructed the spatial correlation functions among quarks and antiquarks produced by $q \bar q$ pair production at mid-rapidity in heavy-light ion collisions.  Depending on the choice of correlation function and length scale, the process may receive contributions from single-pair production \eqref{e:CSsingle} or double-pair production \eqref{e:CSdouble}, with double-pair production opening up many possibilities for entanglement between the pairs.  All of these cases are constructed by tracing the appropriate fermion spin and color flows over combinations of the fundamental building block \eqref{e:buildblock}, the dressed amplitude to produce a single $q \bar q$ pair.

Because of the length and complexity of the full analysis, we have chosen to divide this calculation into parts.  This paper constitues Part I of the analysis, in which we have calculated the single-pair contribution to $\mathcal{C}_{q \bar q}$ \eqref{e:2Dcorr} explicitly and outlined the length scales over which various double-pair contributions occur.  The primary difficulty for the double-pair case arises from the proliferation of the number and complexity of the Wilson line color traces, as illustrated in the specific example \eqref{e:intexample}.  Each double-pair topology admits 16 distinct cases for the interactions due to the various time orderings, and there are 3 distinct topologies associated with fermion entanglement and another 3 with no fermion entanglement.  For this reason, we defer the presentation and analysis of these results for Part II of the analysis in a future publication.  

The calculations presented here are limited to the quasi-classical approximation, in which the Wilson line interactions are rapidity independent.  At high enough energies, small-$x$ evolution will modify this, so the proper inclusion of evolution in the total collision energy $s$ will be a priority among future extensions.  We have focused on constructing the general form of the interactions in terms of Wilson line multipoles, which can then be evaluated in a variety of methods.  For analytic results which can be understood intuitively on an event-averaged basis, we can take advantage of the MV model, supplemented with the large-$N_c$ limit as needed.  For a more detailed and realistic evaluation of our correlation functions, we can instead sample the color multipoles using Monte Carlo techniques~\cite{Dumitru:2011vk}; this approach would be well-suited to studying event-by-event fluctuations.  

Ultimately, the goal of this analysis is to characterize the profile of quark and antiquark correlations in the initial stages of heavy-ion collisions.  A final outcome of this work should be the development of a numerical code which can initialize not only the energy density, but also other conserved charges like flavor and baryon number.  Then, when coupled with state-of-the-art hydrodynamics techniques, we will be able to address novel questions about the role of these conserved charges throughout the evolution of heavy-ion collisions.  One possible observable that may be sensitive to these intial-state correlations is the production of $J/\psi$ and other quarkonium states, because the rates of quarkonium regeneration may well be affected by the presence of initial-state quark pairs in close proximity.  The initial-state (anti)quarks produced with these correlations may also translate, after hydrodynamic evolution and hadronization, into observable correlations between baryons and anti-baryons~\cite{Adam:2016iwf}.  The resulting correlations between same-sign or opposite-sign hadrons may also provide an important conventional background for the Chiral Magnetic Effect~\cite{Kovner:2017gab}.  Although much work still remains to be done to fully assess the potential impact of conserved charges in the initial state of heavy-ion collisions, we believe this analysis represents an important step toward their incorporation in the next generation of hydrodynamic codes.

%
\acknowledgements{The authors wish to thank G. Chirilli, Y. Makris, J. Noronha-Hostler, Y. Kovchegov, and R. Venugopalan for useful discussions. MM especially thanks S. Floerchinger with whom an earlier collaboration~\cite{Floerchinger:2015efa} brought some of the questions addressed in this work.  We thank N.~Armesto for a critical reading of this manuscript and all of the authors of Ref.~\cite{Altinoluk:2016vax} for clarifying the technical details of that work. This work was initiated while the authors were supported by the U.S. Department of Energy, Office of Science, Office of Nuclear Physics under Award Number DE-SC0004286.  At its completion, this work has been supported in part by The U.S. Department of Energy grant DE-FG02-03ER41260 and the BEST (Beam Energy Scan Theory) DOE Topical Collaboration (MM), DOE Contract No. DE-AC52-06NA25396 and the DOE Early Career Program (MS), 
the European Research Council 39 grant HotLHC ERC-2011-StG-279579, Ministerio de Ciencia e Innovaci\'on of Spain under project FPA2014-58293-C2-1-P and Unidad de Excelencia Mar\'ia de Maeztu under project
MDM-2016-0692, Xunta de Galicia (Conseller\'ia de Educaci\'on) and FEDER (DW).}
%

\appendix

%
\section{Derivation of the Wave Functions}
\label{app:WFs} 
\renewcommand{\theequation}{A\arabic{equation}}
  \setcounter{equation}{0}
%

For the early pair production $\Psi_1$ (see Fig.~\ref{f:WF}), there is an instantaneous piece $\Psi_1^{inst}$ and a non-instantaneous piece $\Psi_1^{non}$.  Using the rules and conventions of LFPT as in \cite{Kovchegov:2012mbw} (except for the light-cone metric; they use $\gPM = 2$, but here we keep $\gPM$ arbitrary), we have in the $A^+ = 0$ light-cone gauge
\begin{align}
\Psi_1^{non} &\equiv \frac{1}{p^+} \frac{1}{2 \gpm [ p^- - q^- - (p-q)^- ]} (-g) 
\Big[ \ubar{\sigma_v^\prime}(p-q) \, \slashed{\epsilon}_\lambda^* (q) \, U_{\sigma_v} (p) \Big]
\notag \\ &\times 
\frac{1}{q^+} \frac{1}{2 \gpm [p^- - k^- - (q-k)^- - (p-q)^-]} (-g)
\Big[ \ubar{\sigma'} (k) \, \slashed{\epsilon}_\lambda (q) \, V_\sigma (q-k) \Big]
\notag \\ \notag \\ & =
2 g^2 \delta_{\sigma_v \sigma_v^\prime} 
\frac{\alpha (1-\alpha)}{(\bm{k}-\alpha \bm{q})_T^2 + m^2 + \alpha (1-\alpha) q_T^2}
\notag \\ & \hspace{2cm} \times
\bigg\{
\frac{1}{q^+} \Big[ \ubar{\sigma'} (k) \, \gamma^+ \, V_\sigma (q-k) \Big] -
\frac{\bm{q}_\bot^i}{q_T^2} \Big[ \ubar{\sigma'} (k) \, \bm{\gamma}_\bot^i \, V_\sigma (q-k) \Big]
\bigg\} ,
\end{align}
with $\alpha \equiv \frac{k^+}{q^+}$ and $m$ the mass of the produced (anti)quarks (note that we take the valence quark to be massless, but allow for the possibility of heavy $q \bar q$ production) and a sum over the gluon spin $\lambda$ implied.  The instantaneous part is
\begin{align}
\Psi_1^{inst} &= \frac{1}{p^+} \left( \frac{g^2}{(q^+)^2} 
\Big[ \ubar{\sigma_v^\prime}(p-q) \, \gamma^+ \, U_{\sigma_v} (p) \Big]
\Big[ \ubar{\sigma'}(k) \, \gamma^+ \, V_{\sigma} (q-k) \Big] \right)
\notag \\ & \hspace{2cm} \times
\frac{1}{2 \gpm [p^- - k^- - (q-k)^- - (p-q)^-]}\
\notag \\ \notag \\ & =
-2 g^2 \delta_{\sigma_v \sigma_v^\prime} \frac{\alpha (1-\alpha)}{(\bm{k} - \alpha \bm{q})_T^2 + m^2 + \alpha(1-\alpha) q_T^2} \Big[ \frac{1}{q^+} \ubar{\sigma'} (k) \, \gamma^+ \, V_\sigma (q-k) \Big] .
\end{align}
Thus the instantaneous part cancels the $\gamma^+$ component of the non-instantaneous part, leaving only
\begin{align}
\Psi_1 &= -2 g^2 \delta_{\sigma_v \sigma_v^\prime} \frac{\alpha (1-\alpha)}{(\bm{k} - \alpha \bm{q})_T^2 + m^2 + \alpha(1-\alpha) q_T^2} \: \frac{\bm{q}_\bot^i}{q_T^2} \, \Big[ \ubar{\sigma'} (k) \, \bm{\gamma}_\bot^i \, 
V_\sigma (q-k) \Big] .
\end{align}
The late pair production case $\Psi_3$ (see Fig.~\ref{f:WF}) can be simply obtained from $\Psi_1$:
\begin{subequations}
\begin{align}
\Psi_3^{non} & = \frac{p^- - q^- - (p-q)^-}{q^- - k^- - (q-k)^-} \, \Psi_1^{non} =
\frac{\alpha (1-\alpha) q_T^2}{(\bm{k} - \alpha \bm{q})_T^2 + m^2} \, \Psi_1^{non}
\\
\Psi_3^{inst} &= - \Psi_1^{inst} .
\end{align}
\end{subequations}
Then the sum of the two wave functions is 
\begin{subequations}
\begin{align}
\Psi_1 + \Psi_3 &= \frac{p^- - k^- - (p-q)^- - (q-k)^-}{q^- - k^- - (q-k)^-} \: \Psi_1^{non} = - \Psi_2 ,
\end{align}
\end{subequations}
where $\Psi_2$ is precisely the time ordering with early emission of the gluon and late pair production (see Fig.~\ref{f:WF}).  The fact that $\Psi_1 + \Psi_2 + \Psi_3 = 0$ confirms that the production amplitude is zero in the absence of scattering.

Taking $\Psi_1$ and $\Psi_2$ as the independent wave functions, we have
\begin{subequations}
\begin{align}
\Psi_1 &= - 2 g^2 \frac{\alpha (1-\alpha)}{(\bm{k} - \alpha \bm{q})_T^2 + m^2 + \alpha (1-\alpha) q_T^2}
\: \frac{\bm{q}_\bot^i}{q_T^2} \,  \Big[ \ubar{\sigma'}(k) \, \bm{\gamma}_\bot^i \, V_\sigma (q-k) \Big]
\\ 
\Psi_2 &= -2 g^2 \frac{\alpha (1-\alpha)}{(\bm{k} - \alpha \bm{q})_T^2 + m^2} \bigg\{
\frac{1}{q^+} \Big[ \ubar{\sigma'} (k) \, \gamma^+ \, V_\sigma (q-k) \Big] -
\frac{\bm{q}_\bot^i}{q_T^2} \Big[ \ubar{\sigma'} (k) \, \bm{\gamma}_\bot^i \, V_\sigma (q-k) \Big]
\bigg\} ,
\end{align}
\end{subequations}
where we have dropped $\delta_{\sigma_v \sigma_v^\prime}$ as understood for eikonal gluon emission.  Using the spinor matrix elements
\begin{subequations}
\begin{align}
\frac{1}{q^+} \Big[ \ubar{\sigma'}(k) \, \gamma^+ \, V_\sigma (q-k) \Big] &=
2 \sqrt{\alpha(1-\alpha)} \: \delta_{\sigma \, , \, -\sigma'}
\\
\frac{\bm{q}_\bot^i}{q_T^2} \Big[ \ubar{\sigma'} (k) \, \bm{\gamma}_\bot^i \, V_\sigma (q-k) \Big] &=
\delta_{\sigma \, , \, -\sigma'} \, \frac{1}{\sqrt{\alpha(1-\alpha)}} \, \left[ \alpha + 
(1-2\alpha) \frac{\bm{q} \cdot \bm{k}}{q_T^2} - i \sigma' \, \frac{\bm{q} \times \bm{k}}{q_T^2} \right]
\notag \\ &-
m \sigma' \, \delta_{\sigma \sigma'} \, \frac{1}{\sqrt{\alpha (1-\alpha)}} 
\left[ \frac{\bm{q}_\bot^1}{q_T^2} - i \sigma' \, \frac{\bm{q}_\bot^2}{q_T^2} \right]
\end{align}
\end{subequations}
we obtain the momentum space expressions \eqref{e:WFmom}.

%
\section{Pauli Matrix Algebra}
\label{app:Pauli} 
\renewcommand{\theequation}{B\arabic{equation}}
  \setcounter{equation}{0}
%

The following properties of the Pauli matrices make it straightforward to compute the spin traces when contracting wave functions:
\begin{subequations} \label{e:Pauli}
\begin{align}
\tr[ \mathds 1 ] &= 2 \\
\tr[ \tau_i ] &= 0 \\
\tr[ \tau_i \tau_j ] &= 2 \delta_{i j} \\
\tr[ \tau_i \tau_j \tau_k ] &= 2i \, \epsilon_{i j k} \\
\tr[ \tau_i \tau_j \tau_k \tau_\ell ] &= 2 (\delta_{i j} \delta_{k \ell} - \delta_{i k} \delta_{j \ell} + \delta_{i \ell} \delta_{j k}) \\
(\tau_i)^2 &= 1 \\
\{ \tau_i , \tau_j \} &= 2 \delta_{i j} \\
\tr[ ( [\bm \tau] \times \bm f) ( [\bm \tau] \times \bm g) ] &= 2 \bm f \cdot \bm g \\
\tr[ \tau_3 ( [\bm \tau] \times \bm f) ( [\bm \tau] \times \bm g) ] &= 2 i \bm f \times \bm g \\
\tr[ ( [\bm \tau] \times \bm f) ( [\bm \tau] \times \bm g) ( [\bm \tau] \times \bm h) ( [\bm \tau] \times \bm k) ] &= 
     2 [ \bm f \cdot \bm g \, \bm h \cdot \bm k - \bm f \cdot \bm h \, \bm g \cdot \bm k + \bm f \cdot \bm k \, \bm g \cdot \bm h ] .
\end{align}
\end{subequations}
%

%
\bibliography{baryonfluct}
%

\end{document}